\documentclass[fleqn]{revtex4}

\usepackage{hyperref}
\pdfoutput=1

\usepackage{amsmath}
\usepackage{graphicx}
\usepackage{enumerate}
\usepackage{dcolumn}
\usepackage{longtable}

\begin{document}

\title{Calculation of the relativistic Bethe logarithm in the two-center problem}

\author{Vladimir I. Korobov}
\affiliation{Bogolyubov Laboratory of Theoretical Physics, Joint Institute
for Nuclear Research, Dubna 141980, Russia}
\author{L.~Hilico}
\author{J.-Ph.~Karr}
\affiliation{Laboratoire Kastler Brossel, UPMC-Paris-6, ENS, CNRS\\
Case 74, 4 place Jussieu, 75005 Paris, France}

\begin{abstract}
We present a variational approach to evaluate relativistic corrections of order $\alpha^2$ to the Bethe logarithm for the ground electronic state of the Coulomb two center problem. That allows to estimate the radiative contribution at $m\alpha^7$ order in molecular-like three-body systems such as hydrogen molecular ions $\mbox{H}_2^+$ and $\mbox{HD}^+$, or antiprotonic helium atoms. While we get 10 significant digits for the nonrelativistic Bethe logarithm, calculation of the relativistic corrections is much more involved especially for small values of bond length $R$. We were able to achieve a level of 3-4 significant digits starting from $R=0.2$ bohr, that will allow to reach $10^{-10}$ relative uncertainty on transition frequencies.
\end{abstract}

\maketitle

\section*{Introduction}

Considerable effort is currently devoted to high-precision laser spectroscopy of three-body molecular (or molecule-like) systems such as HD$^+$~\cite{Bre12,Koel12}, H$_2^+$~\cite{Karr12} and antiprotonic helium~\cite{Hori11}. These experiments aim at improving the present accuracy of the electron-to-proton and -antiproton mass ratios, for which the uncertainty of spectroscopic data, as well as of theoretical calculations of transition frequencies, should reach a level of about 0.1 ppb. Systematic evaluation of leading QED corrections up to the $m\alpha^6$ order has improved the theoretical precision in hydrogen molecular ions~\cite{Kor08} and antiprotonic helium~\cite{Kor08b} to 0.3-0.4 ppb and 1 ppb, respectively. The main source of theoretical uncertainty is the $m\alpha^7$ order one-loop self-energy correction \cite{Eides}, which so far has been evaluated only in hydrogenlike systems. Considering the aimed for accuracy, it is enough to calculate the relativistic Bethe logarithm  with 3-4 significant digits and thus it may be obtained in the framework of the adiabatic approximation, i.e.\ for an electron in the field of two fixed nuclei.

For hydrogen-like ions, the one-loop self-energy contribution to the binding energy of an electron is traditionally expressed as follows~\cite{SapYen}
\begin{equation}\label{E_expan}
\begin{array}{@{}l}
\displaystyle
\Delta E_{\rm 1-loop} = \frac{m_e\alpha}{\pi}\frac{(Z\alpha)^4}{n^3}\>
   \Bigl\{
      \Bigl[A_{41}(n)\ln\,[(Z\alpha)^{-2}]+A_{40}(n)\Bigr]+(Z\alpha)A_{50}(n)
\\[3mm]\displaystyle\hspace{20mm}
      +(Z\alpha)^2\Bigl[
         A_{62}(n)\ln^2[(Z\alpha)^{-2}]+A_{61}(n)\ln\,[(Z\alpha)^{-2}]+A_{60}(n)
      \Bigr]
      +\dots
   \Bigr\} \, .
\end{array}
\end{equation}
It is known that among $m\alpha^5$ order terms, the Bethe logarithm (which appears in the low-energy part of the nonlogarithmic contribution $A_{40}$) is the most difficult quantity for numerical evaluation. Similarly, at the $m\alpha^7$ order the low-energy part of $A_{60}$ contains the relativistic Bethe logarithm \cite{Pac92,Jen03,PRL05,JCP05} which gives rise to even more severe difficulties. In the present work, we describe a numerical method which allows to obtain these quantities with very good accuracy for a two-center system.

The paper is organized as follows. In Sec.~\ref{def-bl}, we briefly outline the origin of Bethe logarithm contributions in a nonrelativistic quantum electrodynamics (NRQED) approach \cite{CasLep,Kin96} and give their precise definition. In Sec.~\ref{asympt}, the asymptotic behavior of the integrands is derived. In Sec.~\ref{num} the numerical method is described in detail, and finally, the nonrelativistic and relativistic Bethe logarithms are calculated for the hydrogen atom, hydrogen molecular ions and antiprotonic helium.

\section{Bethe Logarithm: Definitions} \label{def-bl}

In this Section natural relativistic units ($\hbar\!=\!c\!=\!m=\!1$) are used, while starting from Sec.~\ref{asympt} we switch over to the atomic units, which are more suitable for numerical calculations.

As a starting point, we take the nonrelativistic Hamiltonian
\begin{equation}\label{Hamiltonian}
H = \frac{\mathbf{p}^2}{2m}+V,
\qquad
V = -\frac{Z_1\alpha}{r_1}-\frac{Z_2\alpha}{r_2}
\end{equation}
where $r_1$ and $r_2$ are the distances from the electron to nuclei 1 and 2, respectively. The case of $Z_1\!=\!Z_2\!=\!1$ corresponds to the hydrogen molecular ions and $Z_1\!=\!2$, $Z_2\!=\!-1$ to the antiprotonic helium atom.

\subsection{The NRQED one-loop self-energy at the  $m\alpha^5$ order (low photon energy)} \label{nrbl}

The leading order NRQED interaction with the magnetic field is determined by
\[
H_I^{(0)} =
   -\frac{e}{m}\>\mathbf{p}\cdot\mathbf{A}
   -\frac{e}{2m}\>\boldsymbol{\sigma}\!\cdot\mathbf{B}
\]
The first term in this expression is the "dipole" interaction, while the second one is called Fermi's interaction.

\begin{figure}[h]
\begin{center}
\includegraphics[width=30mm]{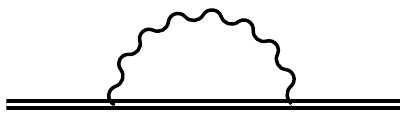}\\
\end{center}
\caption{The NRQED diagram for the leading order one-loop self-energy contribution}\label{NR_SE}
\end{figure}

It may be shown \cite{Pac98} that the NRQED diagram in Fig.~\ref{NR_SE} with the Fermi-type interactions on one or both sides of the transverse photon line give vanishing contributions. Thus the low energy part contribution, which stems from the NRQED diagram in Fig.~\ref{NR_SE} with two dipole vertices may be expressed:
\begin{equation}\label{E_ret}
E_{L-ret} =
   \frac{\alpha^3}{4\pi^2m^2}\int_{|\mathbf{k}|<\Lambda}\frac{d^3k}{k}
   \left(\delta^{ij}-\frac{k^ik^j}{k^2}\right)
   \left\langle
      e^{i\mathbf{kr}_a}\mathbf{p}\left(\frac{1}{E_0-H-k}\right)\mathbf{p}e^{-i\mathbf{kr}_b}
   \right\rangle - \delta m\left\langle\psi_0|\psi_0\right\rangle
\end{equation}
where $\delta m$ is a "mass renormalization" term. Here and throughout this paper it is assumed that in $\bigl\langle\dots\bigr\rangle$ on the left and right-hand sides of brackets stands $\psi_0$, a stationary Schr\"odinger eigenstate of the Hamiltonian operator of Eq.~(\ref{Hamiltonian}), if not otherwise stated.

To get the leading $m\alpha^5$ order contribution, one neglects retardation, replacing the exponential factors in (\ref{E_ret}) by unity, which leads to the nonrelativistic dipole approximation
\begin{equation}\label{E0}
E_{L0} =
   \frac{2\alpha^3}{3\pi m^2}\int_0^\Lambda\,k\,dk\;
   \left\langle
      \mathbf{p}\left(\frac{1}{E_0-H-k}\right)\mathbf{p}
   \right\rangle - \delta m\left\langle\psi_0|\psi_0\right\rangle =
   \frac{2\alpha^3}{3\pi m^2}\int_0^\Lambda\,k\,dk\;P_{nd}(k) - \delta m\left\langle\psi_0|\psi_0\right\rangle
\end{equation}
with
\begin{equation} \label{Pnd}
P_{nd}(k) =
   \left\langle
      \mathbf{p}\left(E_0-H-k\right)^{-1}\mathbf{p}
   \right\rangle.
\end{equation}

The integral in~(\ref{E0}) contains a linearly divergent term which corresponds to the electron's mass renormalization, as was shown by Bethe in 1947~\cite{Bet47}. It also contains a logarithmic term, where the dependance on the cutoff parameter $\Lambda$ is canceled by the logarithmic contribution from the high-energy part \cite{BS}. After these two terms are dropped out, the remaining nonlogarithmic contribution at order $m\alpha^5$ may be written ($E_h=m\alpha^2$ is the Hartree energy)
\begin{equation}\label{numerator_BL}
\mathcal{N}(n;R) =
   \int_0^{E_h}\,k\,dk
   \left\langle
      \mathbf{p}\left(\frac{1}{E_0-H-k}+\frac{1}{k}\right)\mathbf{p}
   \right\rangle
   +\int_{E_h}^\infty\,\frac{dk}{k}\,
   \left\langle
      \mathbf{p}\frac{(E_0\!-\!H)^2}{E_0\!-\!H\!-\!k}\mathbf{p}
   \right\rangle
\end{equation}
and determines the numerator of the Bethe logarithm, while the denominator is expressed by
\begin{equation}\label{denominator_BL}
\mathcal{D}(n;R) = \frac{1}{2}\left\langle \Delta V \right\rangle
\end{equation}
The Bethe logarithm itself is defined as the ratio
\begin{equation}\label{bethe_ratio}
\beta_{nr}(n;R) = \frac{\mathcal{N}}{\mathcal{D}}\>.
\end{equation}
Here $n$ denotes a set of state quantum numbers.

\subsection{One-loop self-energy contributions at the $m\alpha^7$ order}

\begin{figure}[h]
\begin{center}
\includegraphics[width=30mm]{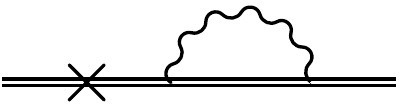}
\includegraphics[width=30mm]{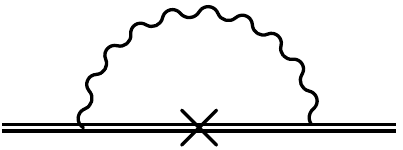}
\includegraphics[width=30mm]{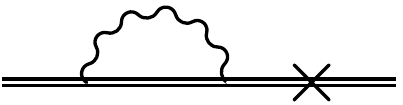}
\end{center}
\caption{NRQED diagrams for relativistic corrections to the electron line, which contribute to self-energy at order $m\alpha^7$.}\label{SE_R1}
\end{figure}

Here, for convenience of reading, we keep the notation of~\cite{JCP05} wherever possible. There are three types of relativistic corrections to the leading order expression~(\ref{E0}), which give a contribution at order $m\alpha^7$ :

\textbf{1.} Relativistic corrections due to the Breit-Pauli interaction (Fig.~\ref{SE_R1})

\begin{equation} \label{E1}
E_{L1} =
   \frac{2\alpha^3}{3\pi m^2}\int_0^\Lambda\,k\,dk\;
   \delta_{H_B}\!\!
   \left\langle
      \mathbf{p}\left(\frac{1}{E_0-H-k}\right)\mathbf{p}
   \right\rangle =
   \frac{2\alpha^3}{3\pi m^2}\int_0^\Lambda\,k\,dk\;P_{rc}^{(1)}(k)
\end{equation}
where
\begin{equation}\label{rel_cor_pert}
\begin{array}{@{}l}
\displaystyle
P_{rc}^{(1)}(k) = \delta_{H_B}\!\!
   \left\langle
      \mathbf{p}\left(\frac{1}{E_0-H-k}\right)\mathbf{p}
   \right\rangle
   \equiv
   2\left\langle
      H_BQ(E_0-H)^{-1}Q\mathbf{p}\left(E_0-H-k\right)^{-1}\mathbf{p}
   \right\rangle
\\[2mm]\hspace{35mm}\displaystyle
   +\left\langle
      \mathbf{p}\left(E_0-H-k\right)^{-1}
      \Bigl(H_B-\left\langle H_B \right\rangle\Bigr)
      \left(E_0-H-k\right)^{-1}\mathbf{p}
   \right\rangle \, .
\end{array}
\end{equation}
Here $Q$ is a projection operator: $Q=I-|\psi_0\rangle\langle\psi_0|$. Eq.~(\ref{rel_cor_pert}) represents the third order term in the Rayleigh-Schr\"odinger perturbation theory. The relativistic Breit-Pauli Hamiltonian for the two-center problem is expressed
\[
H_B = -\frac{\mathbf{p}^4}{8m^3}+\frac{1}{8m^2}\left[4\pi Z_1\delta(\mathbf{r}_1)+4\pi Z_2\delta(\mathbf{r}_2)\right],
\]
the spin interaction is neglected.

For reasons which will be discussed later, it is convenient to split $P^{(1)}_{rc}$ into two parts:
\begin{subequations}
\begin{align}
P^{(1a)}_{rc}(k) &=
   \left\langle
      \mathbf{p}\left(E_0-H-k\right)^{-1}
      \Bigl(H_B-\left\langle H_B \right\rangle\Bigr)
      \left(E_0-H-k\right)^{-1}\mathbf{p}
   \right\rangle
\label{P1a}
\\
P^{(1b)}_{rc}(k) &=
   2\left\langle
      H_BQ(E_0-H)^{-1}Q\mathbf{p}\left(E_0-H-k\right)^{-1}\mathbf{p}
   \right\rangle \, .
\label{P1b}
\end{align}
\end{subequations}

\textbf{2.} Modification of vertex interactions in the self-energy diagram

\begin{figure}[h]
\begin{center}
\includegraphics[width=30mm]{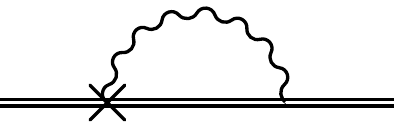}
\includegraphics[width=30mm]{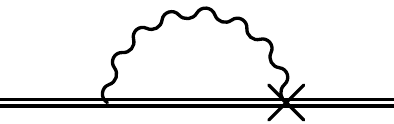}
\end{center}
\caption{The NRQED diagrams for the self-energy with modified vertex interactions at order $m\alpha^7$}\label{SE_R2}
\end{figure}

The next order NRQED interactions ($\sim\!\alpha^2H_I^{(0)}$) with a magnetic field are determined by \cite{Kin96,HFS09}
\[
H_I^{(2)} =
   \frac{e}{2m^3}\,p^2\,\mathbf{p}\cdot\mathbf{A}
   +\frac{e^2}{4m^2}\,\sigma^{ij}(\nabla^jV)A^i,
\]
thus modifying vertex functions as shown on the diagrams in Fig.~(\ref{SE_R2}); one gets
\begin{equation} \label{E3}
E_{L3} =
   \frac{4\alpha^3}{3\pi m^2}\int_0^\Lambda\,k\,dk
   \left\langle
      \delta\mathbf{J}\left(\frac{1}{E_0-H-k}\right)\mathbf{p}
   \right\rangle =
   \frac{4\alpha^3}{3\pi m^2}\int_0^\Lambda\,k\,dk\;P_{rc}^{(2)}(k)
\end{equation}
where $\delta J^i=-p^2p^i-\frac{1}{2}\sigma^{ij}\nabla^jV$, and
\begin{equation} \label{P3}
P_{rc}^{(2)}(k) =
   \left\langle
      \left(-p^2p^i-\frac{1}{2}\sigma^{ij}\nabla^jV\right)
      \left(E_0-H-k\right)^{-1}p^i
   \right\rangle \, .
\end{equation}

\textbf{3.} It remains to consider the effect of retardation (see Eq.~(\ref{E_ret})). We obtain the nonrelativistic quadrupole contribution, which results from the Taylor series expansion of $e^{i(\mathbf{kr})}=1\!+\!i(\mathbf{kr})\!-\!(\mathbf{kr})^2/2\!+\!\dots$

\begin{equation} \label{E2}
E_{L2} =
   \frac{2\alpha^3}{3\pi m^2}\int_0^\Lambda\,k\,dk\;P_{nq}(k)
\end{equation}
\begin{equation} \label{Pnq}
\begin{array}{@{}l}
\displaystyle
P_{nq}(k) =
   \frac{3k^2}{8\pi}\int_S d\Omega_\mathbf{n}
      \left(\delta^{ij}-n^in^j\right)
      \biggl\{
      \left\langle
         p^i(\mathbf{n\cdot r})\left(E_0-H-k\right)^{-1}(\mathbf{n\cdot r})p^i
      \right\rangle
\\[2mm]\hspace{52mm}\displaystyle
      -\left\langle
         p^i(\mathbf{n\cdot r})^2\left(E_0-H-k\right)^{-1}p^i
      \right\rangle
      \biggr\},
\end{array}
\end{equation}
where $\mathbf{k} = k\mathbf{n}$.

Similarly to the nonrelativistic Bethe logarithm considered above, the relativistic Bethe logarithm corresponds to the finite part of the integrals~(\ref{E1}), (\ref{E3}) and (\ref{E2}), i.e. divergent terms in $\Lambda$ must be subtracted~\cite{JCP05}. It is thus essential to study the asymptotic behavior of the integrands in the $k\to\infty$ limit.

\section{Asymptotic behaviour of the integrands at $k\to\infty$} \label{asympt}

 Our approach to obtain asymptotic expansions stems from ideas first formulated by C. Schwartz in~\cite{Schwartz}. The first step is to note that in expression~(\ref{E0}) the integrand's form is that of a second-order perturbation correction. It may be calculated via the first-order perturbation wave function $\psi_1$, which can be obtained by solving the differential equation
\begin{equation}\label{psi1_eq}
(E_0-H-k)\psi_1 = \boldsymbol{\nabla}\psi_0,
\end{equation}
and then one calculates the integrand by evaluating
\begin{equation}
P_{nd}(k) =
   \left\langle
      \psi_0|\boldsymbol{\nabla}|\psi_1
   \right\rangle \, .
\end{equation}

\subsection{Nonrelativistic Bethe logarithm}

The first order nonrelativistic perturbation wave function for $k\to\infty$ to a good extent may be approximated (see \cite{Schwartz,Kor12}) by
\begin{equation}\label{psi1}
\begin{array}{@{}l}\displaystyle
\psi_1(\mathbf{r}) \approx
   \frac{1}{k}
   \left[
      \frac{Z_1\mathbf{r}_1}{r_1}+\frac{Z_2\mathbf{r}_2}{r_2}
   \right]\psi_0(r)
   -\frac{1}{k^2}
   \left\{
      \frac{Z_1\mathbf{r}_1}{r_1^3}
      \left[
         1\!-\!e^{-\mu r_1}\left(1\!+\!\mu r_1\right)
      \right]
      +\frac{Z_2\mathbf{r}_2}{r_2^3}
      \left[
         1\!-\!e^{-\mu r_2}\left(1\!+\!\mu r_2\right)
      \right]
   \right\}\psi_0(r),
\end{array}
\end{equation}
here $\mu=\sqrt{2k}$. This function has a proper behaviour at $r\to0$.

By substituting (\ref{psi1}) into expression
$\frac{1}{k}\left\langle \nabla^2 \right\rangle - \frac{1}{k}
\left\langle \psi_0 |[H,\boldsymbol{\nabla}]| \psi_1 \right\rangle$ (see for details \cite{Schwartz,Kor12})
one gets for the nonrelativistic dipole term
\begin{equation}
\begin{array}{@{}l}\displaystyle
P_{nd}(k) =
   \frac{1}{k}\left\langle\boldsymbol{\nabla}^2\right\rangle
   +\frac{1}{2k^2}\,\left\langle\Delta V\right\rangle
\\[3mm]\displaystyle\hspace{20mm}
   -\frac{1}{k^3}
   \left\{
      \left[Z_1^2\sqrt{2k}\!-\!Z_1^3\ln{k}\right]
         4\pi\left\langle\delta(\mathbf{r}_1)\right\rangle
      +\left[Z_2^2\sqrt{2k}\!-\!Z_2^3\ln{k}\right]
         4\pi\left\langle\delta(\mathbf{r}_2)\right\rangle
   \right\}
   +\dots
\end{array}
\end{equation}
in which all terms (including the last one in $\ln k/k^3$) are correct. For higher-order terms in $1/k$ we use the same expansion as for the hydrogen atom
\begin{equation} \label{asympt-nd}
\sum_{m=1}^M \frac{Q^{nd}_{1m}\sqrt{k}+Q^{nd}_{2m}\ln{k}+Q^{nd}_{3m}}{k^{m+3}},
\end{equation}
which is in the latter case known analytically \cite{Gav70} (See also \cite{Pac92}).

\subsection{Relativistic Bethe logarithm}

Substituting again the wave function $\psi_1$ from (\ref{psi1}) into the matrix elements which appear in the integrands $P_i(k)$ in Eqs.~(\ref{Pnq}), (\ref{P1a}), (\ref{P1b}) and (\ref{P3}), one gets
\begin{subequations}\label{P_as}
\begin{equation}
\begin{array}{@{}l}\displaystyle
P_{nq}(k) =
   -\frac{1}{2}\left\langle \nabla^2 \right\rangle
   -\frac{1}{k}
   \left[
      \frac{\left\langle \nabla^4 \right\rangle}{5}
      +2\pi\left(
         Z_1\left\langle \delta(\mathbf{r}_1) \right\rangle
         \!+\!Z_2\left\langle \delta(\mathbf{r}_2) \right\rangle
      \right)
   \right]
\\[3mm]\displaystyle\hspace{30mm}
   +\frac{2Z_1^2(\sqrt{2k}-Z_1\ln{k})}{k^2}\,
      4\pi\left\langle \delta(\mathbf{r}_1) \right\rangle
   +\frac{2Z_2^2(\sqrt{2k}-Z_2\ln{k})}{k^2}\,
      4\pi\left\langle \delta(\mathbf{r}_2) \right\rangle+\dots
\\[4mm]\displaystyle\hspace{12mm}
 = F_{nq}+\frac{A_{nq}}{k}+\frac{B_{nq}}{k^{3/2}}
   +\frac{C_{nq}\ln{k}}{k^2}+\frac{D_{nq}}{k^2}+\dots
\end{array}
\end{equation}
\begin{equation}
\begin{array}{@{}l}\displaystyle
P_{rc}^{(1a)}(k) =
   -\frac{Z_1^2 \sqrt{2k}}{4k^2}\,4\pi\left\langle\delta(\mathbf{r}_1)\right\rangle
   -\frac{Z_2^2 \sqrt{2k}}{4k^2}\,4\pi\left\langle\delta(\mathbf{r}_2)\right\rangle
   +\dots
\\[3mm]\displaystyle\hspace{12mm}
 = \frac{B_{rc}^{(1a)}}{k^{3/2}}+\frac{D_{rc}^{(1a)}}{k^2}+\dots
\end{array}
\end{equation}
\begin{equation}
\begin{array}{@{}l}\displaystyle
P_{rc}^{(1b)}(k) =
   \frac{2}{k}\left\langle
      \left( H_B\!-\!\left\langle H_B \right\rangle\right)
      (E_0\!-\!H)^{-1}\boldsymbol{\nabla}^2
   \right\rangle
\\[3mm]\displaystyle\hspace{30mm}
   +\frac{Z_1^2(2\sqrt{2k}+ Z_1\ln{k})}{4k^2}\,4\pi\left\langle\delta(\mathbf{r}_1)\right\rangle
   +\frac{Z_2^2(2\sqrt{2k}+ Z_2\ln{k})}{4k^2}\,4\pi\left\langle\delta(\mathbf{r}_2)\right\rangle
   +\dots
\\[3mm]\displaystyle\hspace{12mm}
 = \frac{A_{rc}^{(1b)}}{k}+\frac{B_{rc}^{(1b)}}{k^{3/2}}
   +\frac{C_{rc}^{(1b)}\ln{k}}{k^2}+\frac{D_{rc}^{(1b)}}{k^2}+\dots
\end{array}
\end{equation}
\begin{equation}
\begin{array}{@{}l}\displaystyle
P_{rc}^{(2)}(k) =
   \frac{\left\langle p^4 \right\rangle}{k}
   +\frac{Z_1^2(-\sqrt{8k}+ Z_1\ln{k})}{k^2}\,4\pi\left\langle\delta(\mathbf{r}_1)\right\rangle
   +\frac{Z_2^2(-\sqrt{8k}+ Z_2\ln{k})}{k^2}\,4\pi\left\langle\delta(\mathbf{r}_2)\right\rangle
   +\dots
\\[3mm]\displaystyle\hspace{12mm}
 = \frac{A_{rc}^{(2)}}{k}+\frac{B_{rc}^{(2)}}{k^{3/2}}
   +\frac{C_{rc}^{(2)}\ln{k}}{k^2}+\frac{D_{rc}^{(2)}}{k^2}+\dots
\end{array}
\end{equation}
\end{subequations}

For higher order terms, in the case of $P_{nq}$, $P^{(1a)}_{rc}$ and $P_{rc}^{(2)}$ the form of the asymptotic expansion is found to be similar to the nonrelativistic Bethe logarithm (Eq.~(\ref{asympt-nd})), for example:
\begin{equation}\label{asy1}
P^{(1a)}_{rc}(k) - \frac{B^{(1a)}_{rc}}{k^{3/2}} - \frac{D^{(1a)}_{rc}}{k^2} =
   \sum_{m=1}^M \frac{Q^{(1a)}_{1m}\sqrt{k}+Q^{(1a)}_{2m}\ln{k}+Q^{(1a)}_{3m}}{k^{m+2}},
\end{equation}
with equivalent expressions for $P_{nq}(k)$ and $P_{rc}^{(2)}(k)$. The $P^{(1b)}_{rc}$ term has an essentially different asymptotic behavior:
\begin{equation}\label{asy2}
P^{(1b)}_{rc}(k)
   - \frac{A^{(1b)}_{rc}}{k} - \frac{B^{(1b)}_{rc}}{k^{3/2}}
   - \frac{C^{(1b)}_{rc}\ln{k}}{k^2} - \frac{D^{(1b)}_{rc}}{k^2} =
   \frac{1}{k^2}\sum_{m=1}^M\sum_{n=0}^{m} \frac{S^{(1b)}_{mn}\ln^n{k}}{k^{m/2}}.
\end{equation}
This is one of the reasons why the $P^{(1)}_{rc}$ term has been separated in two contributions.

In actual calculations coefficients of the asymptotic expansion $A$, $B$, $C$, and $F$ are calculated from expectation values of the operators appearing in the Eqs.~(\ref{P_as}a)-(\ref{P_as}d), while the unknown coefficients $D$, $Q$, and $S$ are obtained by fitting of the numerically evaluated integrand using Eqs.~(\ref{asy1})-(\ref{asy2}).

\subsection{Final expression of the relativistic Bethe Logarithm}

In view of the asymptotic expansion obtained in the previous paragraph, the relativistic Bethe logarithm, which is given by the finite part of the integrals~(\ref{E1})-(\ref{E2}), can be written as follows:

\begin{subequations}\label{rel_BL}
\begin{equation}
\mathcal{L} = \beta_1^{(a)}+\beta_1^{(b)}+\beta_2+\beta_3 \, ,
\end{equation}
\begin{equation}
\begin{array}{@{}l}\displaystyle
\beta_1^{(a)} =
   \frac{2}{3}\int_0^{E_h}\,k\,dk
   \left[
      P_{rc}^{(1a)}(k)-\frac{B_{rc}^{(1a)}}{k^{3/2}}
   \right]
\\[3mm]\displaystyle\hspace{20mm}
   +\frac{2}{3}\int_{E_h}^\infty\,k\,dk\,
   \left[
      P_{rc}^{(1a)}(k)-\frac{B_{rc}^{(1a)}}{k^{3/2}}-\frac{D_{rc}^{(1a)}}{k^2}
   \right]
\end{array}
\end{equation}
\begin{equation}
\begin{array}{@{}l}\displaystyle
\beta_1^{(b)} =
   \frac{2}{3}\int_0^{E_h}\,k\,dk
   \left[
      P_{rc}^{(1b)}(k)-\frac{A_{rc}^{(1b)}}{k}-\frac{B_{rc}^{(1b)}}{k^{3/2}}
   \right]
\\[3mm]\displaystyle\hspace{20mm}
   +\frac{2}{3}\int_{E_h}^\infty\,k\,dk\,
   \left[
      P_{rc}^{(1b)}(k)-\frac{A_{rc}^{(1b)}}{k}-\frac{B_{rc}^{(1b)}}{k^{3/2}}
      -\frac{C_{rc}^{(1b)}\ln{k}}{k^2}-\frac{D_{rc}^{(1b)}}{k^2}
   \right]
\end{array}
\end{equation}
\begin{equation}
\begin{array}{@{}l}\displaystyle
\beta_2 =
   \frac{4}{3}\int_0^{E_h}\,k\,dk
   \left[
      P_{nq}(k)-F_{nq}-\frac{A_{nq}}{k}-\frac{B_{nq}}{k^{3/2}}
   \right]
\\[3mm]\displaystyle\hspace{20mm}
   +\frac{4}{3}\int_{E_h}^\infty\,k\,dk\,
   \left[
      P_{nq}(k)-F_{nq}-\frac{A_{nq}}{k}-\frac{B_{nq}}{k^{3/2}}
      -\frac{C_{nq}\ln{k}}{k^2}-\frac{D_{nq}}{k^2}
   \right]
\end{array}
\end{equation}
\begin{equation}
\begin{array}{@{}l}\displaystyle
\beta_3 =
   \frac{2}{3}\int_0^{E_h}\,k\,dk
   \left[
      P_{rc}^{(2)}(k)-\frac{A_{rc}^{(2)}}{k}-\frac{B_{rc}^{(2)}}{k^{3/2}}
   \right]
\\[3mm]\displaystyle\hspace{20mm}
   +\frac{2}{3}\int_{E_h}^\infty\,k\,dk\,
   \left[
      P_{rc}^{(2)}(k)-\frac{A_{rc}^{(2)}}{k}-\frac{B_{rc}^{(2)}}{k^{3/2}}
      -\frac{C_{rc}^{(2)}\ln{k}}{k^2}-\frac{D_{rc}^{(2)}}{k^2}
   \right] \, .
\end{array}
\end{equation}
\end{subequations}
Here the terms which are subtracted from the first line of each equation appear in the expansions due to a formal Taylor series expansion in powers of $(Z\alpha)^2$ of the QED one-loop self-energy correction, which is expressed \cite{SapYen}:
\begin{equation}\label{SE_Dirac}
\begin{array}{@{}l}\displaystyle
\Delta E = -i\frac{e^2}{(2\pi)^4}\int\frac{d^4k}{k^2\!+\!i\epsilon}
     \left\langle
        \overline{\psi}^D_0(\mathbf{r})\left|
           e^{i\mathbf{kr}}\gamma_\mu
           S_F(\mathbf{r},\mathbf{r}',E_0\!-\!k^0)
           \gamma^\mu e^{-i\mathbf{kr}'}
        \right|\psi_0^D(\mathbf{r}')
     \right\rangle
     -\delta m\left\langle\overline{\psi}_0^D|\psi_0^D\right\rangle,
\end{array}
\end{equation}
here $S_F(\mathbf{r},\mathbf{r}',E)$ is the Dirac-Coulomb propagator and $\psi_0^D$ is the Dirac wave function. These extra terms do not appear in asymptotic expansion of the integrand in Eq.~(\ref{SE_Dirac}) and should be withdrawn.

Thus our definition coincides with that of \cite{JCP05} (see prescriptions after Eq.~(3.16) in that Ref.).

\section{Numerical scheme} \label{num}

Here we present a numerical scheme to evaluate the Bethe logarithm~(\ref{bethe_ratio}) and its relativistic corrections~(\ref{rel_BL}) for the two-center Coulomb problem.

\subsection{Variational expansion}

A variational expansion for the electronic wave function is taken in the form ($Z_1\ne Z_2$) \cite{Tso06}:
\begin{equation}\label{exp}
   \Psi_m(\mathbf{r}) = e^{im\varphi}r^{|m|}\sum^{\infty}_{i=1}
 C_{i}e^{-\alpha_{i} r_1 - \beta_{i} r_2},
\end{equation}
where $r$ is a distance from the electron to the $z$-axis and
\[
r=\frac{1}{2R}\sqrt{2r_1^2r_2^2+2r_1^2R^2+2r_2^2R^2-r_1^4-r_2^4-R^4}.
\]
For $Z_1=Z_2$ the variational wave function should be symmetrized
\begin{equation}\label{expsym}
   \Psi(\mathbf{r_{1},r_{2}}) = e^{im\varphi}r^{|m|}\sum^{\infty}_{i=1}
 C_{i}(e^{-\alpha_{i} r_{1} - \beta_{i} r_{2}}\pm
       e^{-\beta_{i} r_{1} - \alpha_{i} r_{2}} ),
\end{equation}
where $(+)$ is used to get a {\em gerade} electronic state and $(-)$ is for an {\em ungerade} state, respectively. Parameters $\alpha_{i}$ and $\beta_{i}$ are generated in a quasi-random manner~\cite{Kor00}
\begin{equation}\label{generator}
  \alpha_{i}=
  \left\lfloor\frac{1}{2}i(i+1)\sqrt{p_{\alpha}}\right\rfloor
                              (A_{2}-A_{1}) + A_{1}.
\end{equation}
Here $\lfloor x\rfloor$ designates the fractional part of $x$, $p_{\alpha}$ is a prime number, and $[A_{1},A_{2}]$ is a real variational interval, which has to be optimized. Parameters $\beta_{i}$ are obtained in a similar way. Details may be found elsewhere \cite{Tso06,Kor07}.

All the integrands in Eqs.~(\ref{Pnd}), (\ref{Pnq}), (\ref{P1a}), (\ref{P1b}) and (\ref{P3}) have the form of a second-order perturbation expression, i.e. they involve an operator $\left(E_0-H-k\right)^{-1}$. We thus diagonalize the matrix of the Hamiltonian for intermediate states to get a set of (pseudo)state wavefunctions $\psi_m$ and energies $E_m$. A similar approach was used to compute the nonrelativistic Bethe logarithm for the three-body Coulomb problem in~\cite{Kor12}.

The basis for intermediate states is constructed as follows.
\begin{enumerate}[(1)]
\item We use a regular basis set (with regular values of the exponents $\alpha,\beta$), similar to that used for the initial state.
\item We build a special basis set with exponentially growing parameters for $r_1$:
\begin{equation}
\left\{
\begin{array}{@{}l}
A_1^{(0)} = A_1,\\[1mm]
A_1^{(n)} = \tau^n A_1,
\end{array}
\right.
\qquad
\left.
\begin{array}{@{}l}
A_2^{(0)} = A_2,\\[1mm]
A_2^{(n)} = \tau^n A_2,
\end{array}
\right.
\end{equation}
where $\tau = A_2/A_1$. Typically $[A_{1},A_{2}] = [2.5,4.5]$, and $n_{max} = 5-7$, which corresponds to the photon energy interval $k \in [0,10^4]$.
\item We add a similar basis set for $r_2$. Note that this last step may be omitted in the case $Z_1 = Z_2$ where the basis is symmetrized.
\end{enumerate}

\subsection{Nonrelativistic Bethe logarithm}

\begin{figure}[t]
\begin{center}$
\begin{array}{cc}
\includegraphics[width=8cm]{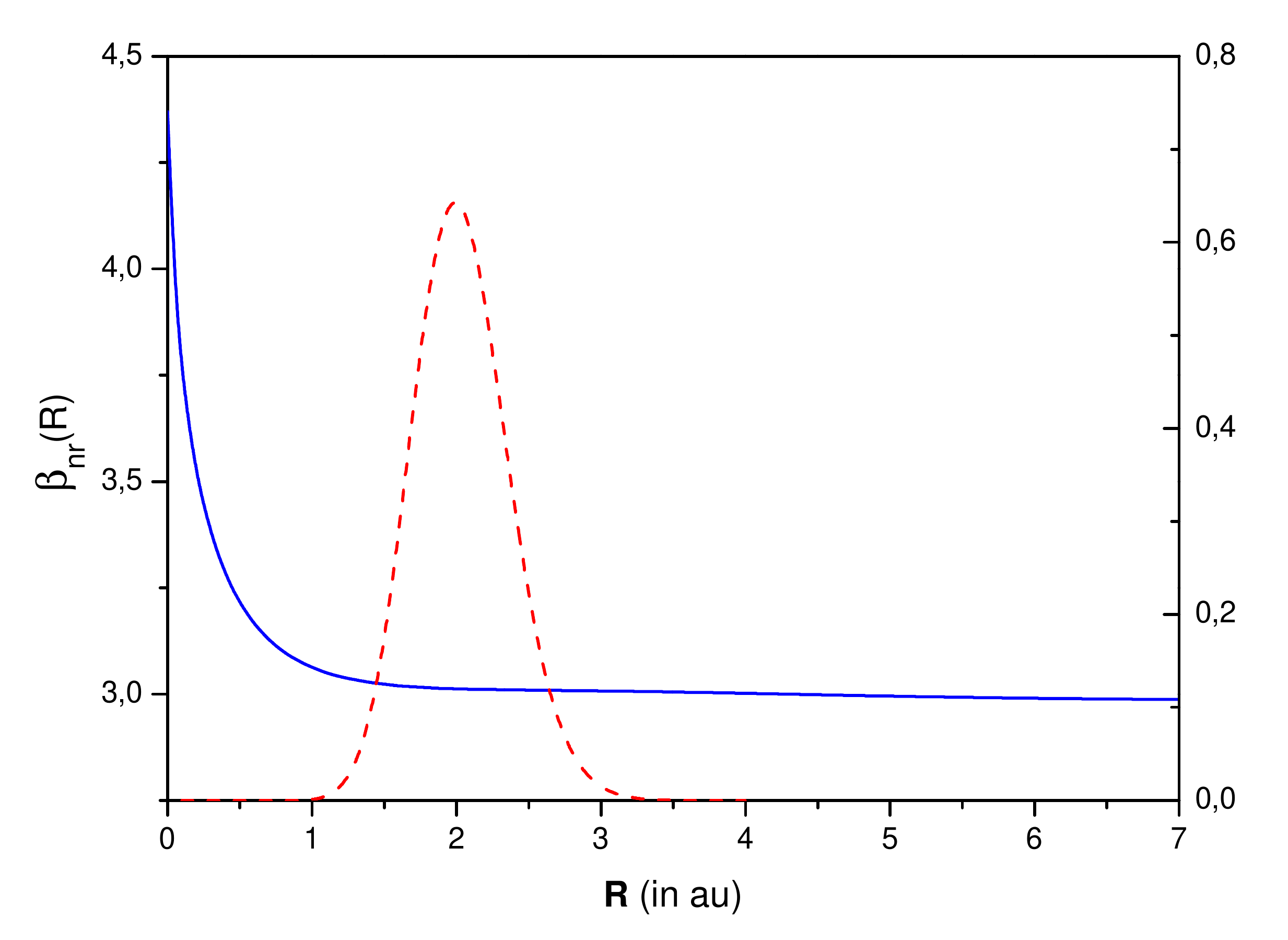} & \includegraphics[width=8cm]{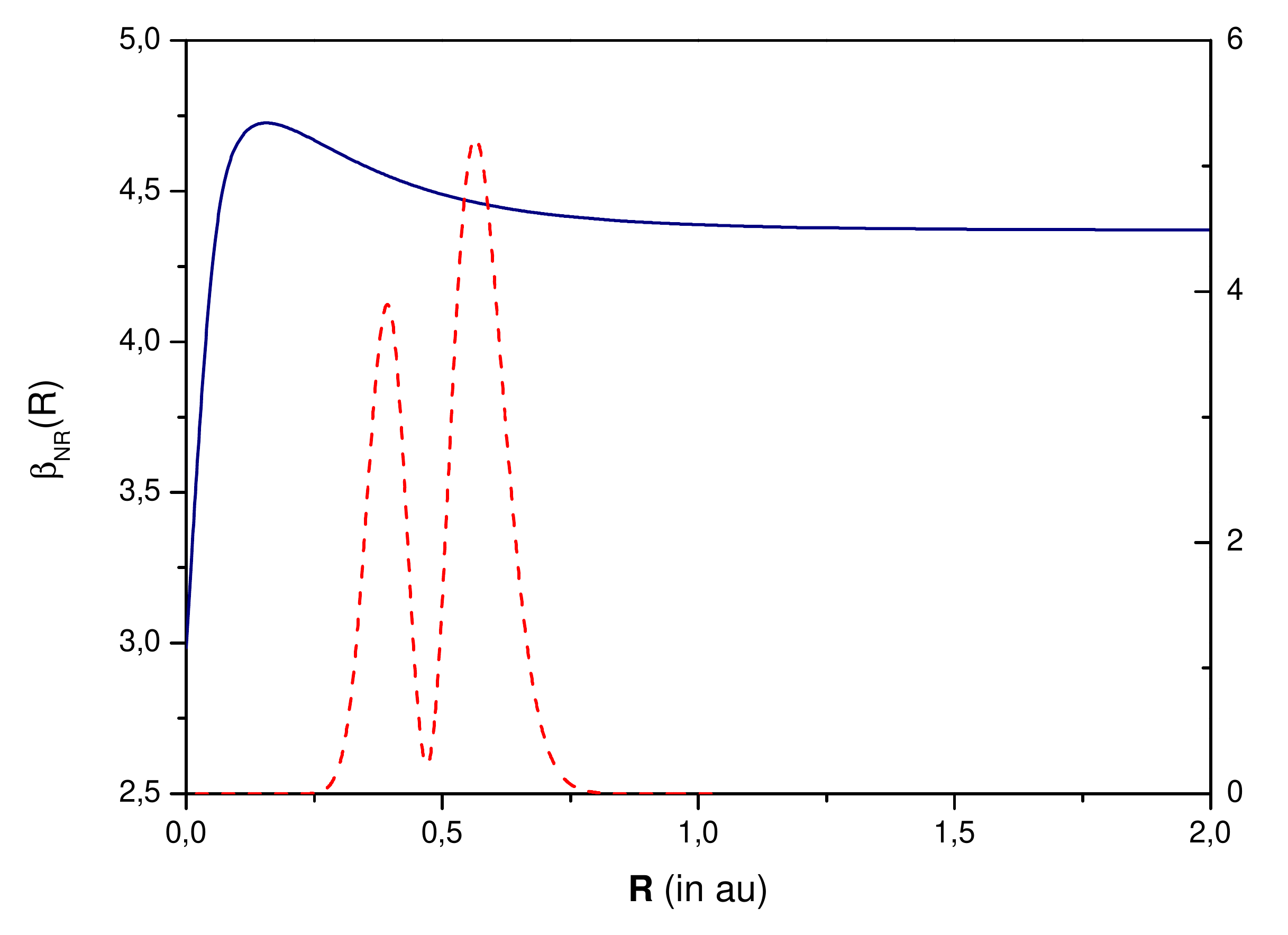}
\end{array}$
\end{center}
\caption{The nonrelativistic Bethe logarithm for the ground ($1s\sigma$) electronic state for $Z_1 = Z_2 = 1$ (left) and $Z_1 = 2$, $Z_2 = -1$ (right). Dashed lines show the vibrational wave functions for the ground state in H$_2^+$ ion (left) and the $(36,34)$ state in $^4$He$^+\bar{p}$ (right).} \label{Bethe-nr-fig}
\end{figure}

After expansion on the basis for intermediate state, the expression of the integrand becomes
\begin{equation}
P_{nd}(k) = \sum_m \frac{ \left\langle \psi_0 \right| \boldsymbol{\nabla} \left| \psi_m \right\rangle^2}{E_0 - E_m - k},
\end{equation}
so the integral appearing in the low-energy part of the numerator is
\begin{equation}
\int_0^{\Lambda} k dk P_{nd}(k) = \sum_m \left\langle \psi_0 \right| \boldsymbol{\nabla} \left| \psi_m \right\rangle^2 \left[ \Lambda - \left( E_0 - E_m \right) \ln{\left| \frac{E_0 - E_m}{E_0 - E_m - \Lambda}\right|} \, \right] .
\end{equation}
It remains to calculate the matrix elements of the impulse operator. Its standard components are
\begin{equation}
\begin{array}{@{}l}
\displaystyle
\nabla_0^{(1)} = \nabla_z
\\[3mm]\displaystyle
\nabla_{\pm 1}^{(1)} = \mp \frac{1}{\sqrt{2}} \left( \nabla_x \pm i \nabla_y  \right)
\end{array}
\end{equation}
Assuming, from now on, that $\psi_0$ is a $\sigma$-state, action on $\psi_0$ of the impulse operator may be expressed as follows
\begin{equation}
\begin{array}{@{}l}
\displaystyle
\nabla_0^{(1)} \psi_0 =
   \left[
      \left(z+\frac{R}{2}\right)\frac{1}{r_1}\partial_{r_1}
      +\left(z-\frac{R}{2}\right)\frac{1}{r_2}\partial_{r_2}
   \right] \psi_0,
\\[3mm]\displaystyle
\nabla_{\pm 1}^{(1)} \psi_0 =
   r e^{\pm i \varphi}
   \left(
      \frac{1}{r_1}\partial_{r_1} + \frac{1}{r_2}\partial_{r_2}
   \right)  \psi_0.
\end{array}
\end{equation}
Here $z=(r_1^2\!-\!r_2^2)/(2R)$. Using these relations then calculation of the matrix elements is straightforward \cite{Tso06,Kor07}.

\subsection{Relativistic corrections to the Bethe logarithm}

\begin{figure}[t]
\begin{center}$
\begin{array}{cc}
\includegraphics[width=7cm]{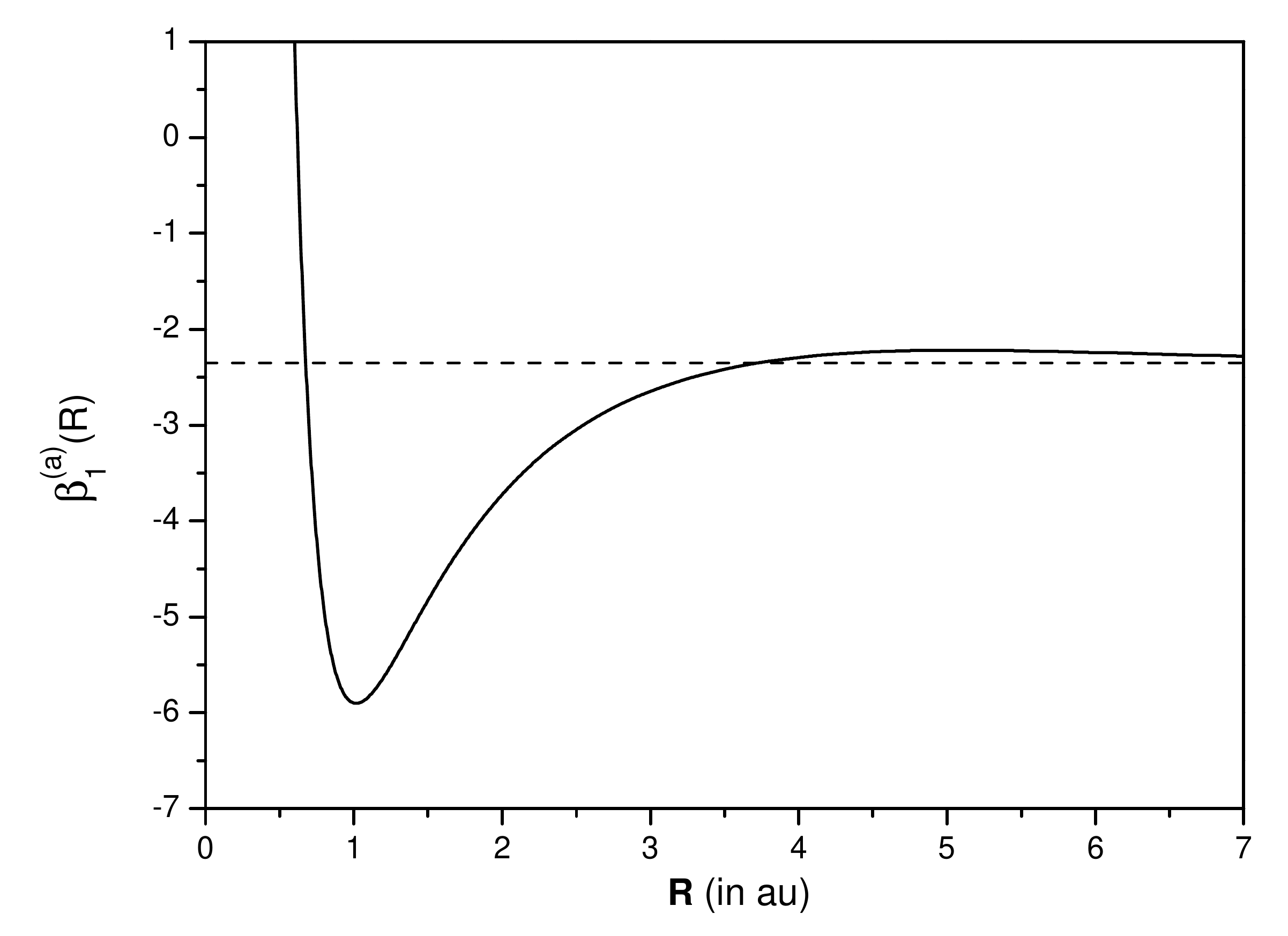} & \includegraphics[width=7cm]{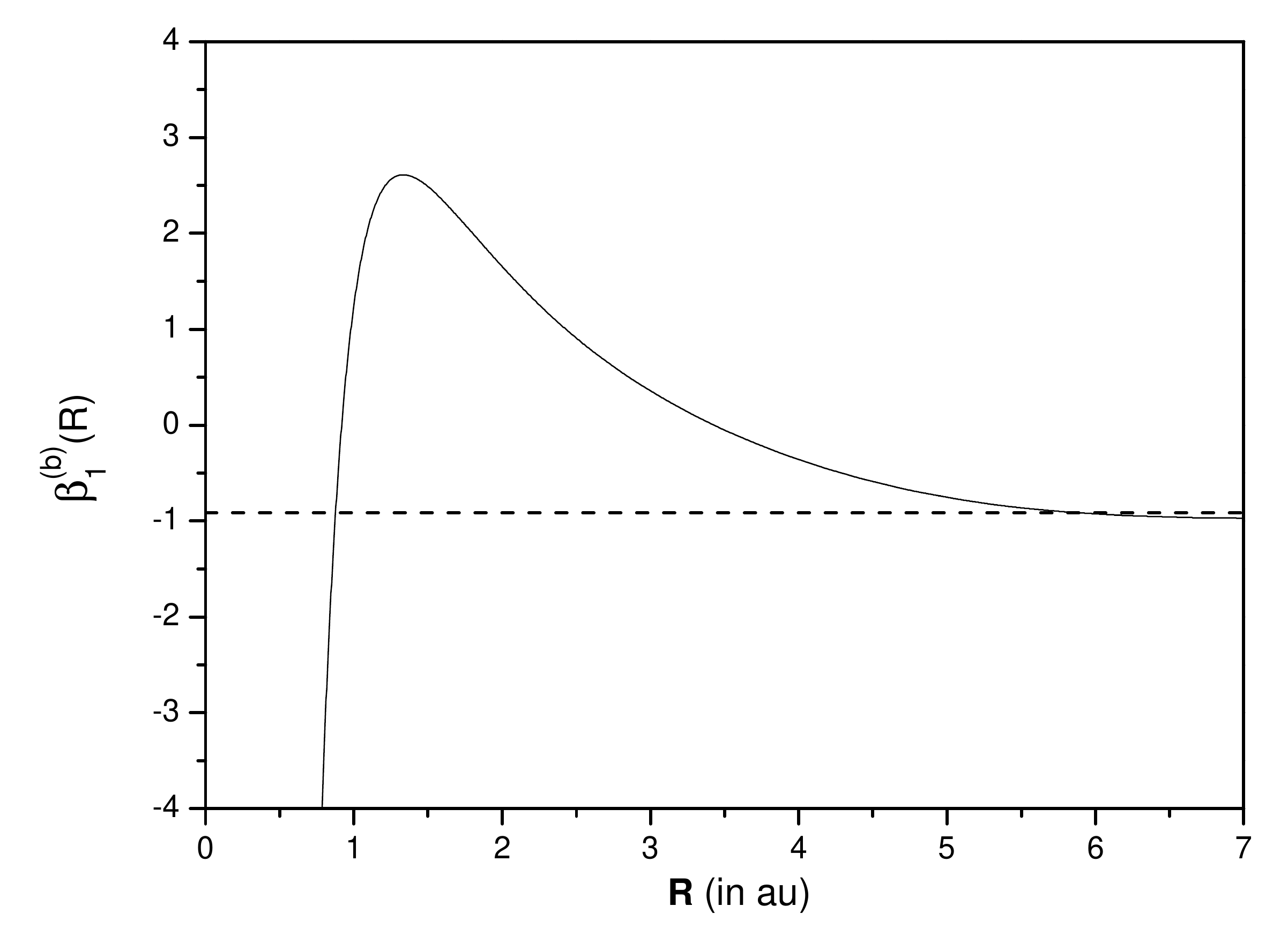} \\
\includegraphics[width=7cm]{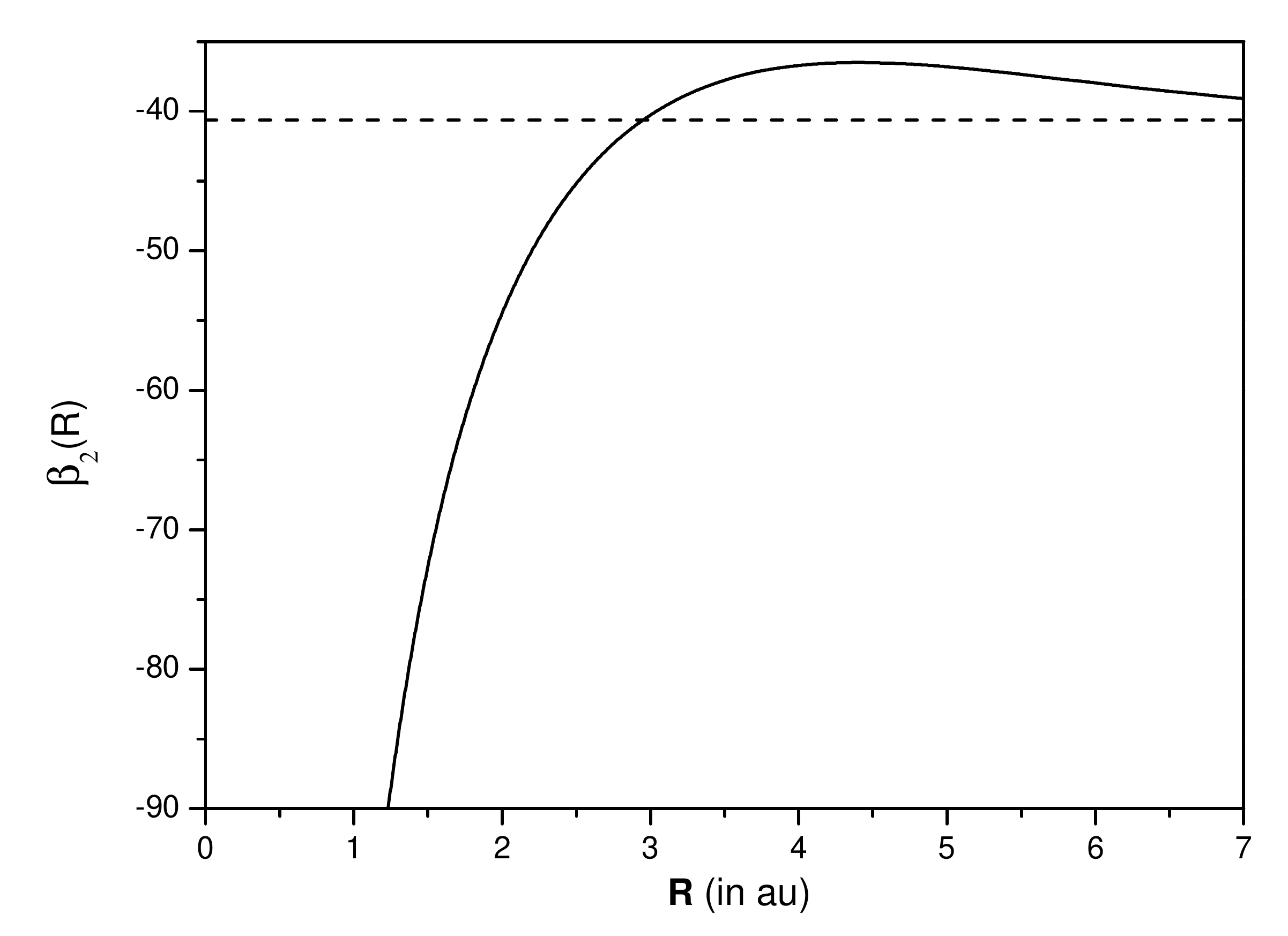} & \includegraphics[width=7cm]{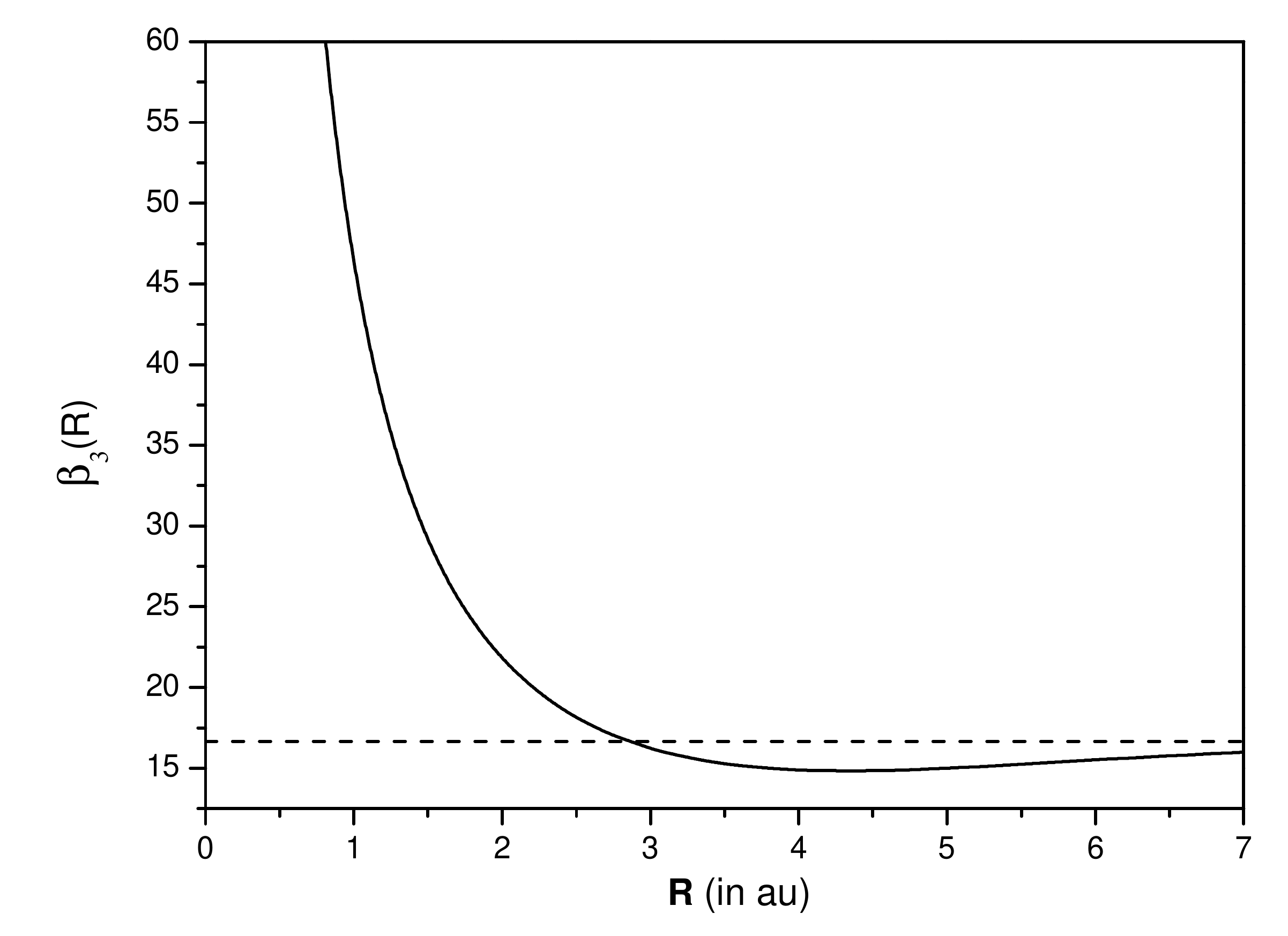}
\end{array}$
\end{center}
\caption{The four contributions to the relativistic Bethe logarithm for the ground ($1s\sigma_g$) electronic state, for $Z_1 = Z_2 = 1$. Dashed lines show their respective values for the 1s state of a hydrogen atom \cite{JCP05}.} \label{beta-h2p-fig}
\end{figure}

We devote one paragraph to each term of Eq.~(\ref{rel_BL}), giving additional details which are necessary for their numerical evaluation. Note that the terms $\beta_1^{(a)}$ and $\beta_1^{(b)}$ are treated numerically in an independent way, which is a further reason for separating these two contributions.

{\bf 1.} For $\beta_1^{(a)}$, the integrand can be written as
\begin{equation}
 P_{rc}^{(1a)}(k) = \left\langle \psi_1 | \left( H_B - \left\langle H_B \right\rangle \right) | \psi_1 \right\rangle.
 \end{equation}
To evaluate $\left\langle\psi_1|\mathbf{p}^4|\psi_1\right\rangle$, it is convenient to use the following identity (see Eq.~(\ref{psi1_eq}))
\begin{equation}
\mathbf{p}^2\psi_1 = 2\big[(E_0-V-k)\psi_1-\nabla\psi_0\big].
\end{equation}
Then
\begin{equation}
\left\langle\psi_1|\mathbf{p}^4|\psi_1\right\rangle =
   4\big\langle
      \psi_1|(E_0\!-\!V\!-\!k)^2|\psi_1
   \big\rangle
   -4\big\langle
      \psi_1|(E_0\!-\!V\!-\!k)|\psi_0
   \big\rangle
   +\left\langle\nabla^2\right\rangle
\end{equation}
and, for arbitrary $k$, $\psi_1(k)$ may be expressed:
\begin{equation}
\psi_1(k) = \sum_m
    \frac{|\psi_m \rangle\langle\psi_m|\nabla|\psi_0\rangle}
    {E_0-E_m-k}.
\end{equation}

{\bf 2.} In order to get $P^{(1b)}_{rc}(k)$ we first solve the equation
\begin{equation}
(E_0-H)\psi_B = \Bigl( H_B-\left\langle H_B \right\rangle \Bigr)\psi_0
\end{equation}
It can be shown that $\psi_B$ behaves at small $r_1$ (or $r_2$) as
\begin{equation}\label{psiB}
\psi_B(r_1,r_2) =
   \left(\frac{Z_1}{4r_1}-\frac{Z_1^2}{2}\ln{r_1}\right)\psi_0(r_1,r_2)
   +\left(\frac{Z_2}{4r_2}-\frac{Z_2^2}{2}\ln{r_2}\right)\psi_0(r_1,r_2)
   +\tilde{\psi}_B(r_1,r_2)
\end{equation}
where $\tilde{\psi}_B(r_1,r_2)$ is a regular function. Then equation for $\tilde{\psi}_B(r_1,r_2)$ may be written
\begin{equation}
(E_0-H)\,\tilde{\psi}_B =
   \Bigl( H_B-\left\langle H_B \right\rangle \Bigr)\psi_0
   +\left[H,\left(\frac{Z_1}{4r_1}+\frac{Z_2}{4r_2}
      -\frac{Z_1^2}{2}\ln{r_1}-\frac{Z_2^2}{2}\ln{r_2}\right)\right]\psi_0
\end{equation}
Thus substituting $\psi_B(r_1,r_2)$ from Eq.~(\ref{psiB}) into Eq.~(\ref{P1b}), one gets
\begin{equation}
\begin{array}{@{}l}
\displaystyle
P^{(1b)}_{rc}(k) =
   2\left\langle
      \left(
         \frac{Z_1}{4r_1}+\frac{Z_2}{4r_2}
         -\frac{Z_1^2}{2}\ln{r_1}-\frac{Z_2^2}{2}\ln{r_2}
      \right)
      Q\mathbf{p}\left(E_0-H-k\right)^{-1}\mathbf{p}
   \right\rangle
\\[4mm]\displaystyle\hspace{25mm}
   +2\left\langle
      \tilde{\psi}_B\bigl|\,Q\mathbf{p}\left(E_0-H-k\right)^{-1}\mathbf{p}\bigr|\,\psi_0
   \right\rangle
\end{array}
\end{equation}
The derivation of matrix elements involving logarithms may be found in the Appendix.

\begin{figure}[t]
\begin{center}
\includegraphics[width=8cm]{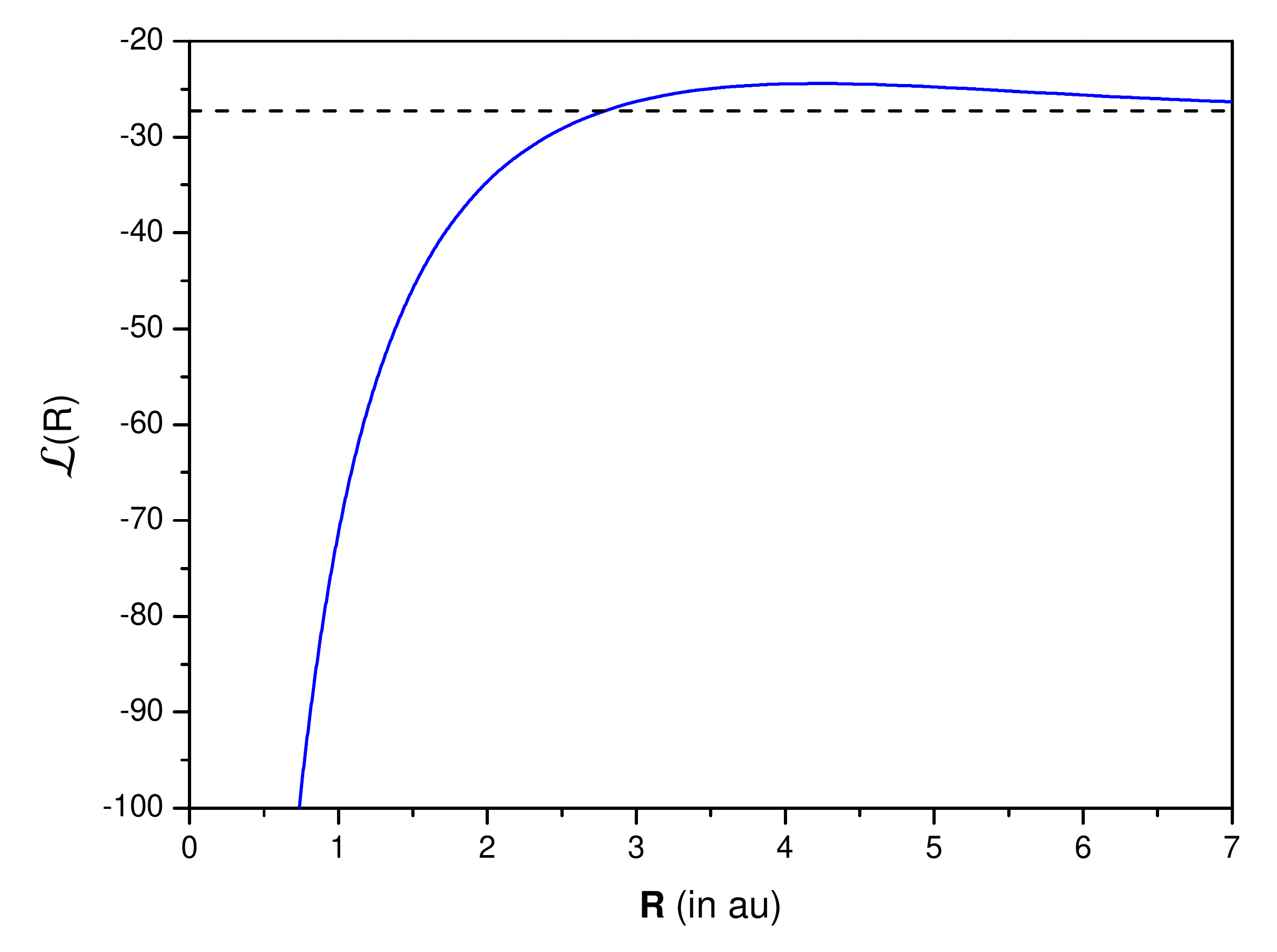}\hspace{5mm}\includegraphics[width=8cm]{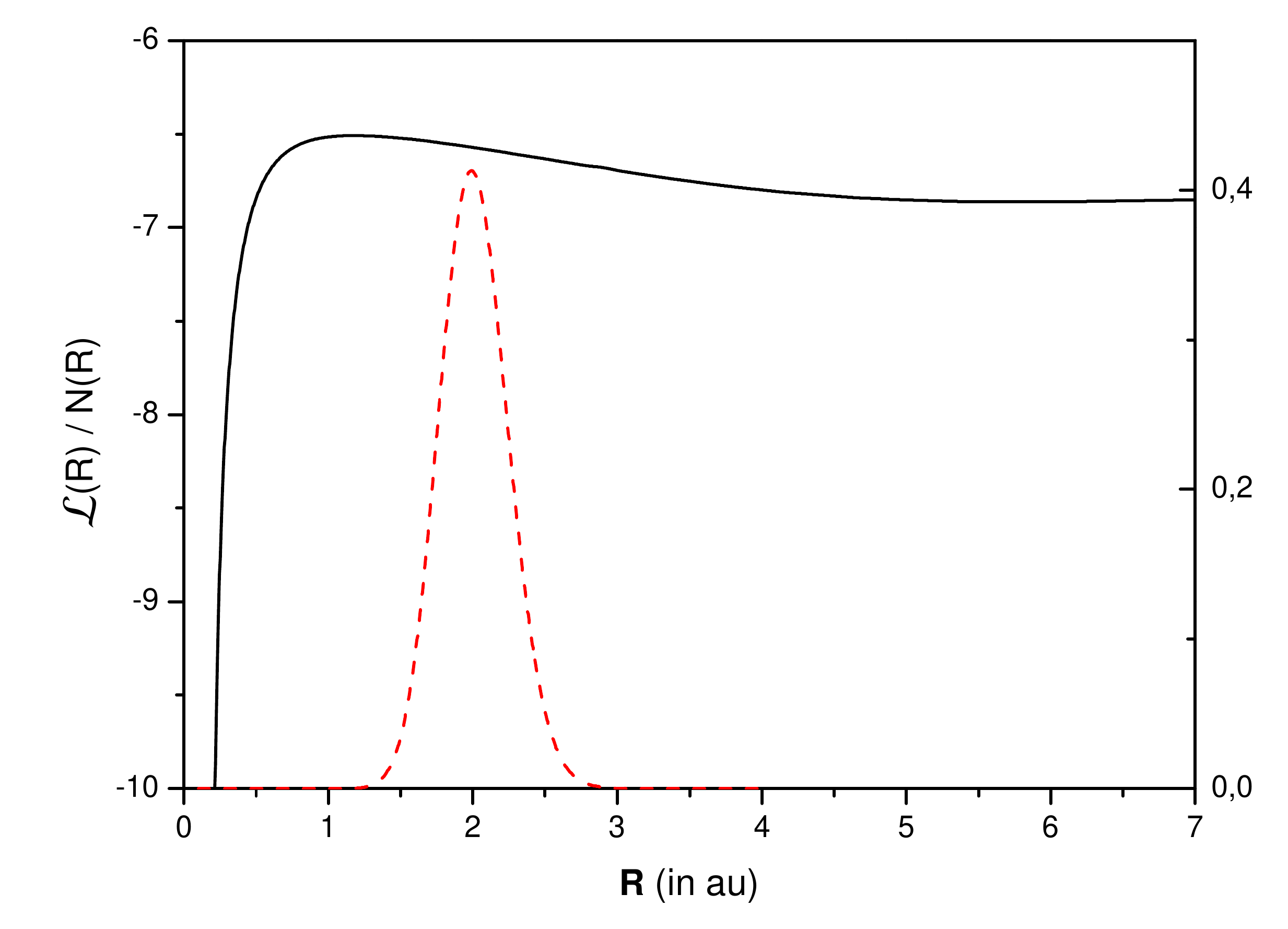}
\end{center}
\caption{(Left) The relativistic Bethe logarithm for the ground ($1s\sigma_g$) electronic state, for $Z_1 = Z_2 = 1$. The dashed line is the $\mathcal{L}(1s)$ for the 1s state of a hydrogen atom \cite{JCP05}. (Right) The same data, but normalized by the delta function distribution: $N(R)=4\pi\left(Z_1^3\delta(\mathbf{r}_1)+Z_2^3\delta(\mathbf{r}_2)\right)$. The dashed line is the vibrational wave function for the ground state of $\mbox{H}_2^+$. \label{rbl-h2p-fig}}
\end{figure}

\begin{table}[t]
\begin{center}
\begin{tabular}{@{\hspace{5mm}}c@{\hspace{1mm}}d@{\hspace{1mm}}d@{\hspace{4mm}}
               d@{\hspace{1mm}}d@{\hspace{5mm}}d@{\hspace{2mm}}d}
\hline\hline
\vrule width0pt height10pt depth4pt
$R$ & \multicolumn{1}{c}{$\beta_1^{(a)}$} & \multicolumn{1}{c}{$\beta_1^{(b)}$} & \multicolumn{1}{c}{$\beta_2$} &
      \multicolumn{1}{c}{$\beta_3$} & \mathcal{L}  & \multicolumn{1}{c}{$\mathcal{L}/N(R)$} \\
\hline
\vrule width0pt height10pt depth4pt
0.2 &  199.124 & -362.792 & -718.74  & 500.9   & -381.5   & -9.4774  \\
0.3 &  71.425  & -155.864 & -408.41  & 244.09  & -248.76  & -7.7149  \\
0.4 &  26.866  & -73.891  & -293.55  & 153.814 & -186.76  & -7.0981  \\
0.5 &  8.5247  & -36.4368 & -232.469 & 111.218 & -149.163 & -6.8191  \\
0.6 &  0.3419  & -17.8754 & -192.985 & 86.9905 & -123.529 & -6.6763  \\
0.7 & -3.4173  & -8.19603 & -164.670 & 71.3995 & -104.884 & -6.5980  \\
0.8 & -5.1044  & -2.99381 & -143.164 & 60.5041 & -90.7581 & -6.5534  \\
0.9 & -5.7669  & -0.16341 & -126.255 & 52.4484 & -79.7371 & -6.5273  \\
1.0 & -5.94144 &  1.36042 & -112.653 & 46.2540 & -70.9802 & -6.5145  \\
1.1 & -5.85242 &  2.14699 & -101.528 & 41.3527 & -63.8807 & -6.5084  \\
1.2 & -5.64166 &  2.50981 & -92.3095 & 37.3908 & -58.0506 & -6.5071  \\
1.3 & -5.37875 &  2.62712 & -84.5938 & 34.1346 & -53.2109 & -6.5097  \\
1.4 & -5.10067 &  2.60269 & -78.0750 & 31.4227 & -49.1503 & -6.5145  \\
1.5 & -4.82677 &  2.49792 & -72.5270 & 29.1400 & -45.7159 & -6.5212  \\
1.6 & -4.56684 &  2.34961 & -67.7744 & 27.2013 & -42.7903 & -6.5294  \\
1.7 & -4.32531 &  2.17994 & -63.6799 & 25.5425 & -40.2828 & -6.5387  \\
1.8 & -4.10371 &  2.00222 & -60.1351 & 24.1145 & -38.1221 & -6.5489  \\
1.9 & -3.90197 &  1.82441 & -57.0531 & 22.8786 & -36.2520 & -6.5598  \\
2.0 & -3.71921 &  1.65108 & -54.3637 & 21.8045 & -34.6274 & -6.5712  \\
2.1 & -3.55413 &  1.48479 & -52.0099 & 20.8676 & -33.2117 & -6.5831  \\
2.2 & -3.40532 &  1.32679 & -49.9446 & 20.0480 & -31.9752 & -6.5953  \\
2.3 & -3.27133 &  1.17755 & -48.1290 & 19.3294 & -30.8933 & -6.6077  \\
2.4 & -3.15078 &  1.03708 & -46.5306 & 18.6985 & -29.9458 & -6.6203  \\
2.5 & -3.04237 &  0.90509 & -45.1223 & 18.1441 & -29.1155 & -6.6329  \\
2.6 & -2.94493 &  0.78117 & -43.8811 & 17.6566 & -28.3883 & -6.6456  \\
2.7 & -2.85740 &  0.66481 & -42.7875 & 17.2283 & -27.7518 & -6.6583  \\
2.8 & -2.77881 &  0.55550 & -41.8248 & 16.8522 & -27.1958 & -6.6709  \\
2.9 & -2.70829 &  0.45272 & -40.9786 & 16.5527 & -26.6814 & -6.6758  \\
3.0 & -2.64510 &  0.35594 & -40.2365 & 16.2347 & -26.2909 & -6.6955  \\
3.2 & -2.53796 &  0.17881 & -39.0229 & 15.7664 & -25.6156 & -6.7192  \\
3.4 & -2.45271 &  0.02095 & -38.1126 & 15.4188 & -25.1256 & -6.7416  \\
3.6 & -2.38557 & -0.12013 & -37.4498 & 15.1695 & -24.7860 & -6.7624  \\
3.8 & -2.33346 & -0.24634 & -36.9899 & 15.0008 & -24.5689 & -6.7815  \\
4.0 & -2.29385 & -0.35913 & -36.6963 & 14.8980 & -24.4513 & -6.7985  \\
4.2 & -2.26461 & -0.45961 & -36.5387 & 14.8490 & -24.4139 & -6.8135  \\
4.4 & -2.24395 & -0.54861 & -36.4913 & 14.8435 & -24.4404 & -6.8264  \\
4.6 & -2.23034 & -0.62682 & -36.5320 & 14.8726 & -24.5165 & -6.8370  \\
4.8 & -2.22246 & -0.69487 & -36.6417 & 14.9288 & -24.6303 & -6.8456  \\
5.0 & -2.21917 & -0.75336 & -36.8039 & 15.0053 & -24.7712 & -6.8523  \\
\hline\hline
\end{tabular}
\end{center}
\caption{Relativistic Bethe logarithm for the ground ($1s\sigma_g$) electronic state, for $Z_1 = Z_2 = 1$.}\label{beta-h2p}
\end{table}

{\bf 3.} For the evaluation of $P_{nq}(k)$, Eq.~(\ref{Pnq}) must be transformed to separate contributions from operators of different ranks. Denoting $A_s = \left( E_0 - H - k \right)^{-1}$,
\begin{equation}
P_{nq}(k) =
\frac{3}{8\pi}\int d\Omega_{\mathbf{n}}
   \left(\delta^{ij}-n^in^j\right)n^ln^m\int k^3dk
   \left\{ \left\langle
      p^ir^l\,A_s\,r^mp^j
   \right\rangle
   - \left\langle
      p^ir^lr^m\,A_s\,p^j
   \right\rangle \right\}
\end{equation}
and one obtains
\begin{equation}
\begin{array}{@{}l}
\displaystyle
\frac{3}{8\pi}\int d\Omega_{\mathbf{n}}
   \left(\delta^{ij}-n^in^j\right)n^ln^m\int k^3dk
   \left\langle
      p^ir^l\,A_s\,r^mp^j
   \right\rangle
\\[3mm]\hspace{15mm}\displaystyle
 = \int k^3dk
   \left\{
      \frac{3}{10}
      \left\langle
         \left[S^{(2)}_{ij}\right]^\dagger A_s\,S^{(2)}_{ij}
      \right\rangle
      -\frac{1}{4}
      \Bigl\langle
         \left[\mathbf{p\!\times\!r}\right]A_s\left[\mathbf{r\!\times\!p}\right]
      \Bigr\rangle
   \right\}
\\[3mm]\hspace{15mm}\displaystyle
 = \int k^3dk
   \left\{
      \frac{9}{20}
      \left\langle
         \left[S^{(2)}_{\mu}\right]^\dagger A_s\,S^{(2)}_{\mu}
      \right\rangle
      -\frac{1}{4}
      \Bigl\langle
         \left[\mathbf{p\!\times\!r}\right]A_s\left[\mathbf{r\!\times\!p}\right]
      \Bigr\rangle
   \right\}
\end{array}
\end{equation}
\begin{equation}
\begin{array}{@{}l}
\displaystyle
\frac{3}{8\pi}\int d\Omega_{\mathbf{n}}
   \left(\delta^{ij}-n^in^j\right)n^ln^m\int k^3dk
   \left\langle
      p^ir^lr^m\,A_s\,p^j
   \right\rangle
\\[3mm]\hspace{15mm}\displaystyle
 = \int k^3dk
   \left\{
      \frac{2}{5}
      \left\langle
         p^i\,r^2A_s\,p^i
      \right\rangle
      -\frac{1}{5}
         \left\langle
            (\mathbf{pr})r^j\,A_s\,p^j
         \right\rangle
   \right\}.
\end{array}
\end{equation}
Here the quadrupole operator is defined as
\begin{equation}
S^{(2)}_{ij} =
   \frac{1}{2}\left[r_ip_j+r_jp_i-\frac{2}{3}\delta_{ij}
      \left(\mathbf{r\cdot p}\right)\right]
\end{equation}
and its standard components are
\begin{equation}
\begin{array}{@{}l}
\displaystyle
S_0^{(2)} = S_{zz}^{(2)} = \frac{1}{3} \left( 2 z p_z - x p_x - y p_y \right)
\\[3mm]\displaystyle
S_{\pm 1}^{(2)} =
   \mp \sqrt{\frac{2}{3}} \left( S_{xz}^{(2)} \pm i S_{yz}^{(2)} \right) =
   \mp \frac{1}{\sqrt{6}} \left[ x p_z + z p_x \pm i\left( y p_z + z p_y \right) \right]
\\[3mm]\displaystyle
S_{\pm 2}^{(2)} =
   \frac{1}{\sqrt{6}} \left( S_{xx}^{(2)} - S_{yy}^{(2)} \pm 2i S_{xy}^{(2)} \right) =
   \frac{1}{\sqrt{6}} \left[ x p_x - y p_y \pm i \left( x p_y + y p_x \right) \right]
\end{array}
\end{equation}
Matrix elements can be computed from the relations (valid for a $\sigma$ state):
\begin{equation}
\begin{array}{@{}l}
\displaystyle
S_0^{(2)} =
   \frac{1}{3}\left( 2 z p_z\! +\! r_+ p_- \!+\! r_- p_+ \right) =
   -\frac{i}{3}\left[
      2z\left[
         \left(z+\frac{R}{2}\right)\frac{1}{r_1}\partial_{r_1}
         +\left(z-\frac{R}{2}\right)\frac{1}{r_2}\partial_{r_2}
      \right]
      -r^2\left(
         \frac{1}{r_1}\partial_{r_1}\!+\!\frac{1}{r_2}\partial_{r_2}
      \right)
   \right],
\\[4mm]\displaystyle
S_{\pm1}^{(2)} =
   \frac{1}{\sqrt{3}}\left(r_{\pm}p_z+zp_{\pm}\right) =
   \pm\frac{i}{\sqrt{6}}\,r e^{\pm i \varphi}
   \left(
       \left(2z+\frac{R}{2}\right)\frac{1}{r_1}\partial_{r_1}
       +\left(2z-\frac{R}{2}\right)\frac{1}{r_2}\partial_{r_2},
   \right)
\\[4mm]\displaystyle
S_{\pm2}^{(2)} =
   \frac{2}{\sqrt{6}}\,r_\pm p_\pm =
   -\frac{i}{\sqrt{6}}\,r^2e^{\pm i2\phi}
   \left(\frac{1}{r_1}\partial_{r_1}+\frac{1}{r_2}\partial_{r_2}\right).
\end{array}
\end{equation}
Here $r_\pm$ and $p_\pm$ are standard components of vector operators $\mathbf{r}$ and $\mathbf{p}$, respectively.

Finally, the matrix elements of the $\left[\mathbf{r}\times\mathbf{p}\right]$ operator can be obtained from
\begin{equation}
\left[\mathbf{r}\times\mathbf{p}\right]_{\pm1} =
   \frac{ire^{\pm i\phi}}{\sqrt{2}}\frac{R}{2}
   \left(\frac{1}{r_1}\partial_{r_1}-\frac{1}{r_2}\partial_{r_2}\right)
\end{equation}

\begin{figure}[t]
\begin{center}
\includegraphics[width=8cm]{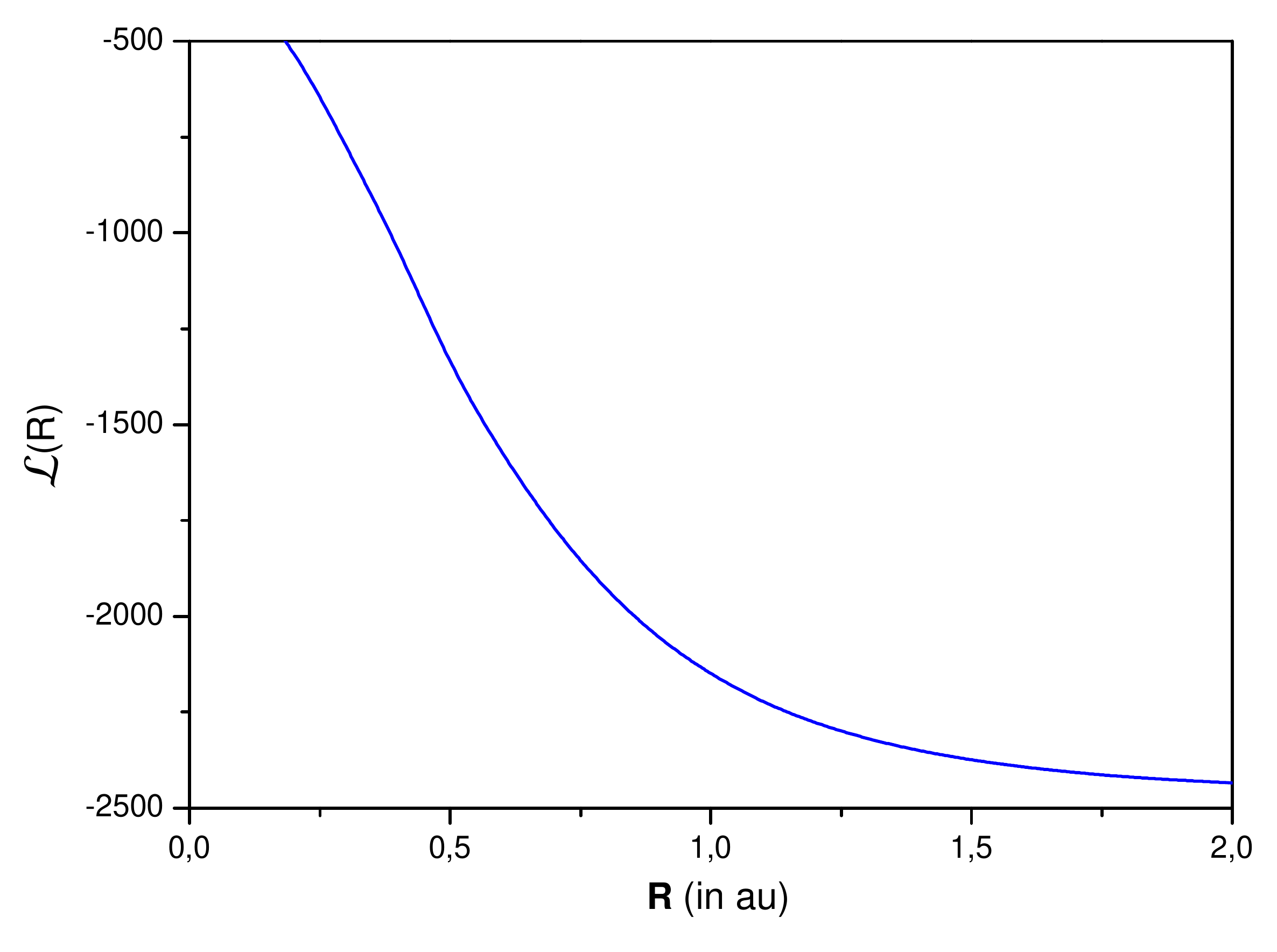}\hspace{5mm}
\includegraphics[width=8cm]{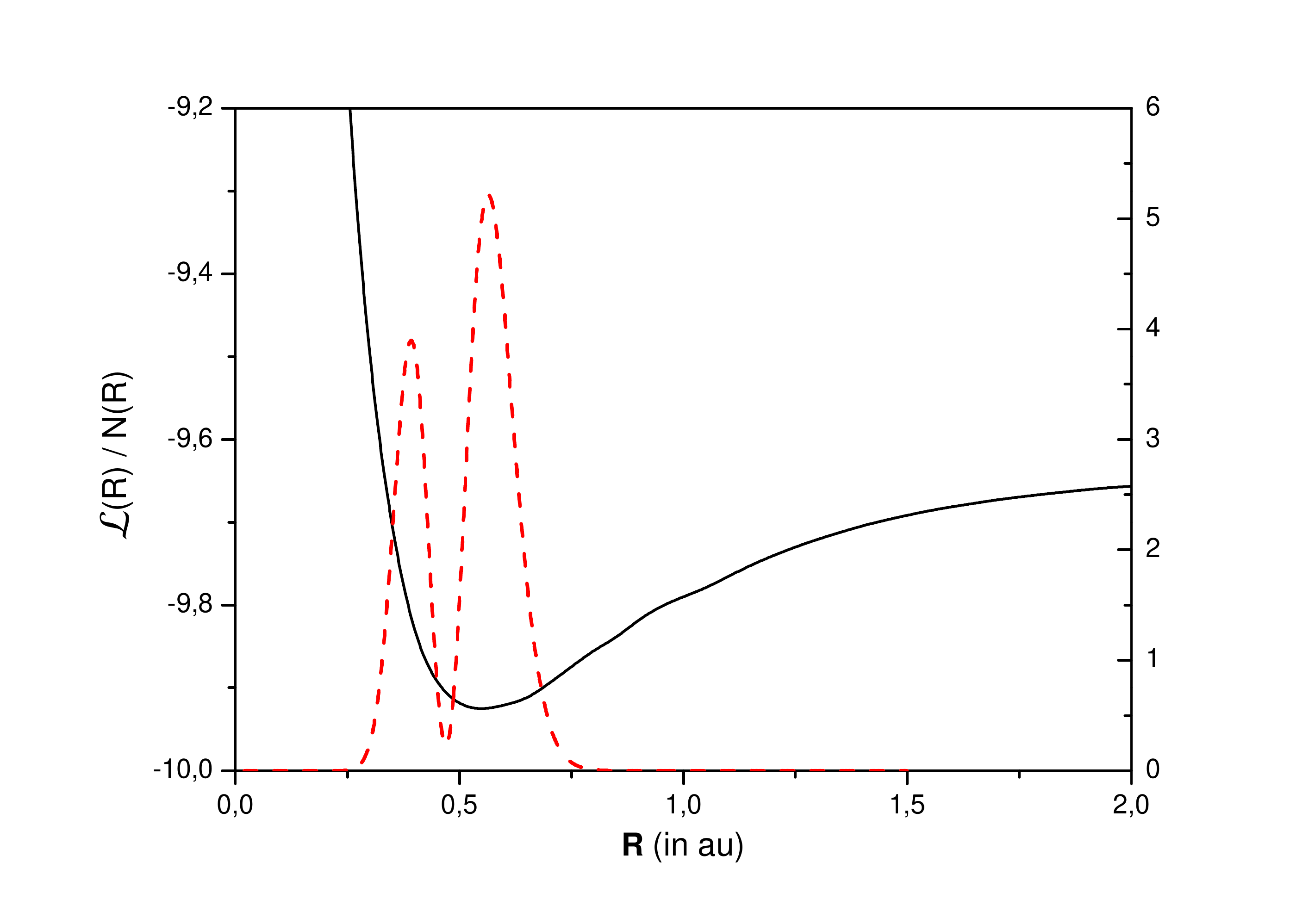}
\end{center}
\caption{(Left) The relativistic Bethe logarithm for the ground ($1s\sigma$) electronic state for $Z_1 = 2$ and $Z_2 = -1$. (Right) The same data, but normalized by the delta function distribution: $N(R)=4\pi\left(Z_1^3\delta(\mathbf{r}_1)+Z_2^3\delta(\mathbf{r}_2)\right)$. The dashed line is the vibrational wave function for the $(36,34)$ state of $^4\mbox{He}^+\bar{p}$.} \label{rbl-hep-fig}
\end{figure}

{\bf 4.}
In the expression of $P_{rc}^{(2)}(k)$ in Eq.~(\ref{P3}), the two terms can be respectively written as
\begin{equation}
   -2\left\langle
      \left(E_0-V\right)p^i
      \left(E_0-H-k\right)^{-1}p^i
   \right\rangle
\end{equation}
and
\begin{equation}
   \frac{1}{2}\left\langle
      \left(
         Z_1\frac{[\boldsymbol{\sigma}\times\mathbf{r}_1]^i}{r_1^3}
         +Z_2\frac{[\boldsymbol{\sigma}\times\mathbf{r}_2]^i}{r_2^3}
      \right)
      \left(E_0-H-k\right)^{-1}p^i
   \right\rangle.
\end{equation}
We now show that the second term does not contribute here. Since right-hand transition does not change spin, then on the left-hand side we may keep only those terms, which contain $\sigma_z$:
\begin{equation}
   \mp\frac{i}{2}
   \left\langle
      \sigma_z \frac{r e^{\pm i\varphi}}{\sqrt{2}}
      \left(
         \frac{Z_1}{r_1^3}
         +\frac{Z_2}{r_2^3}
      \right)
      \left(E_0-H-k\right)^{-1}p^i
   \right\rangle
\end{equation}
This contribution results in fine splitting of the main line, which is spin-dependent, and thus will not be considered here.

\begin{table}[t]
\begin{center}
\begin{tabular}{@{\hspace{2mm}}c@{\hspace{5mm}}d@{\hspace{6mm}}d@{\hspace{1mm}}d@{\hspace{5mm}}d@{\hspace{1mm}}d}
\hline\hline
\vrule width0pt height10pt depth4pt
$R$ & \multicolumn{1}{c}{$\beta_1$} & \multicolumn{1}{c}{$\beta_2$} & \multicolumn{1}{c}{$\beta_3$} &
      \mathcal{L}  & \multicolumn{1}{c}{$\mathcal{L}/N(R)$} \\
\hline
\vrule width0pt height10pt depth4pt
0.20  &    29.65  &   -584.1  &   62.514 &  -491.9 &  -8.647  \\
0.25  &    17.558 &   -799.9  &  159.373 &  -622.9 &  -9.1887 \\
0.30  &     1.741 &   -998.9  &  236.611 &  -760.5 &  -9.5198 \\
0.35  &   -15.627 &  -1191.9  &  305.525 &  -902.0 &  -9.7143 \\
0.40  &   -33.158 &  -1382.7  &  369.822 & -1046.0 &  -9.8402 \\
0.45  &   -49.903 &  -1566.8  &  430.590 & -1186.1 &  -9.8973 \\
0.50  &   -65.195 &  -1743.3  &  487.854 & -1320.6 &  -9.9213 \\
0.55  &   -78.929 &  -1909.9  &  541.306 & -1447.5 &  -9.9272 \\
0.60  &   -91.000 &  -2064.5  &  590.636 & -1564.8 &  -9.9209 \\
0.65  &  -101.172 &  -2207.3  &  635.664 & -1672.8 &  -9.9137 \\
0.70  &  -109.370 &  -2335.7  &  676.345 & -1768.7 &  -9.8946 \\
0.75  &  -116.075 &  -2451.0  &  712.787 & -1854.2 &  -9.8745 \\
0.80  &  -121.547 &  -2553.7  &  745.209 & -1930.0 &  -9.8546 \\
0.85  &  -125.908 &  -2645.8  &  773.901 & -1997.8 &  -9.8408 \\
0.90  &  -127.846 &  -2727.0  &  799.189 & -2055.6 &  -9.8199 \\
0.95  &  -130.235 &  -2797.5  &  821.408 & -2106.3 &  -9.8013 \\
1.00  &  -131.770 &  -2860.8  &  840.902 & -2151.6 &  -9.7898 \\
1.10  &  -133.766 &  -2963.8  &  872.905 & -2224.6 &  -9.7653 \\
1.20  &  -134.563 &  -3041.6  &  897.365 & -2278.7 &  -9.7402 \\
1.30  &  -134.679 &  -3101.3  &  916.090 & -2319.8 &  -9.7204 \\
1.40  &  -134.430 &  -3147.1  &  930.468 & -2351.0 &  -9.7042 \\
1.50  &  -134.006 &  -3182.5  &  941.556 & -2374.9 &  -9.6915 \\
1.60  &  -133.517 &  -3209.9  &  950.148 & -2393.2 &  -9.6810 \\
1.70  &  -133.025 &  -3231.3  &  956.847 & -2407.4 &  -9.6727 \\
1.80  &  -132.561 &  -3248.2  &  962.102 & -2418.6 &  -9.6664 \\
1.90  &  -132.140 &  -3261.4  &  966.255 & -2427.2 &  -9.6607 \\
2.00  &  -131.766 &  -3272.0  &  969.562 & -2434.2 &  -9.6565 \\
\hline\hline
\end{tabular}
\end{center}
\caption{Relativistic Bethe logarithm for the ground ($1s\sigma$) electronic state, for $Z_1 = 2$ and $Z_2 = -1$.}\label{beta-hep}
\end{table}

\section{Results}

\subsection{Nonrelativistic Bethe logarithm}

First we perform calculations of the nonrelativistic Bethe logarithm for the two-center Coulomb problem for two cases: $Z_1=Z_2=1$ and $Z_1=2$, $Z_2=-1$. Results are plotted in Fig.~\ref{Bethe-nr-fig}. The numerical data for $\beta_{nr}(R)$ is as accurate as 10 significant digits and has been obtained for a large range of internuclear bond lengths $R=0-7$ a.u. with a step of 0.05 a.u. For the hydrogen molecular ion our results are in a good agreement with previous calculations \cite{Kolos}. The complete Tables are too lengthy to be reported here and may be found in the Supplemental Material \cite{Supp}.

\subsection{Relativistic corrections to the Bethe logarithm}

The numerical scheme has been tested on the ground state of the hydrogen atom. Our results are
\[
\begin{array}{@{}l}
\beta_1(1s) = -3.268\,213\,19,
\\[1mm]
\beta_2(1s) = -40.647\,026\,693.
\\[1mm]
\beta_3(1s) = ~~16.655\,330\,436,
\end{array}
\]
They are in perfect agreement with the ones of \cite{JCP05}.

The main results of this work are presented in Tables~\ref{beta-h2p} and~\ref{beta-hep}. Particular behaviour of different components of the relativistic Bethe logarithm $\mathcal{L}$, namely $\beta_1^{(a)}$, $\beta_1^{(b)}$, $\beta_2$, $\beta_3$, for a case of a hydrogen molecular ion are shown on Fig.~\ref{beta-h2p-fig}. On Figures~\ref{rbl-h2p-fig} and \ref{rbl-hep-fig} our final results for the low energy contribution $\mathcal{L}(R)$ are plotted.

The numerical calculations have been performed in multiple precision arithmetic with 64 decimal digits. Special care has been required for the nonrelativistic quadrupole contribution at small $R$, in this case a 96 decimal digit arithmetic has been used.

It is worth noting that in the region where the wave function is essentially nonzero the quantity $\mathcal{L}/N(R)$ is about constant: $\mathcal{L}/N(R)\approx6.5$ for $Z_1=Z_2=1$ and $\mathcal{L}/N(R)\approx9.7$ for $Z_1=2$, $Z_2=-1$. That may help in a qualitative estimate within 2-3 digit accuracy of the one-loop self-energy contribution to the transition frequencies of not only hydrogen molecular ions but also of the neutral hydrogen molecule $\mbox{H}_2$.

In conclusion, we have computed the low-energy part of the $m\alpha^7$ order self-energy correction for the two center problem, with a numerical accuracy that exceeds 3 significant digits, in the whole range of internuclear distances $R\in[0.2,7]$. Calculation of the high energy part is under consideration now. A complete result then should be averaged over vibrational wave functions to get proper correction for ro-vibrational transition frequencies at the $m\alpha^7$ order. Taking into account as well the vacuum polarization and two-loop electron self-energy contributions at this order we expect that a relative precision for transition frequencies will be at a level of $10^{-10}$ or better both for hydrogen molecular ions and for antiprotonic helium.

\section{Acknowledgements}

The authors are in debt to Krzysztof Pachucki for helpful discussions. V.I.K. acknowledges support of the Russian Foundation for Basic Research under Grant No.~12-02-00417-a. This work was supported by \'Ecole Normale Sup\'erieure, which is gratefully acknowledged.

\appendix

\section*{Appendix: Two-center matrix elements involving $\ln{r_1}$ or $\ln{r_2}$}

In this Appendix we want to consider an analytical evaluation of the two-center integrals of type \cite{Tso06}, which contain either $\ln{r_1}$ or $\ln{r_2}$ as a multiplier. We start from the basic integral:
\begin{equation}
\Gamma_{\ln{r_1},00}(\alpha,\beta) =
   \int dr_1dr_2\,
   \Bigl[
      \ln{r_1}\cdot e^{-\alpha r_1 - \beta r_2}
   \Bigr]
\end{equation}

Using identities
\[
\int_R^\infty dr\>\ln{r}\>e^{-\gamma r} =
   \frac{1}{\gamma}
   \left[
      \mbox{E}_1\!\bigl(R\gamma\bigr)+\ln{R}\cdot e^{-\gamma R}
   \right]
\]
and
\[
\int_0^R dr\>\ln{r}\>e^{-\gamma r} =
   -\frac{1}{\gamma}
   \left[
      \mbox{E}_1\!\bigl(R\gamma\bigr)+\ln{\gamma}+
      \ln{R}\cdot e^{-\gamma R}+\gamma_E
   \right]
\]
one may get
\begin{subequations}
\begin{equation}
\begin{array}{@{}l}
\displaystyle
\int_R^\infty dr_1\,\ln{r_1}\,e^{-\alpha r_1}
\int_{r_1-R}^{r_1+R}dr_2\,e^{-\beta r_2} =
\\[4mm]\displaystyle\hspace{20mm}
=\frac{1}{\beta}\int_R^\infty dr_1
   \left[
      \ln{r_1}\,e^{-\alpha r_1}
      \left(
         e^{-\beta(r_1-R)}-e^{-\beta(r_1+R)}
      \right)
   \right]
\\[4mm]\displaystyle\hspace{20mm}
=\frac{e^{\beta R}-e^{-\beta R}}{\beta}\times
    \frac{\mbox{E}_1\!\bigl(R(\alpha\!+\!\beta)\bigr)+\ln{R}\cdot
          e^{-(\alpha+\beta)R}}{\alpha+\beta}
\\[3mm]
\end{array}
\end{equation}
and
\begin{equation}
\begin{array}{@{}l}
\displaystyle
\int_0^R dr_1\,\ln{r_1}\,e^{-\alpha r_1}
   \int_{R-r_1}^{R+r_1}dr_2\,e^{-\beta r_2} =
\\[4mm]\displaystyle\hspace{20mm}
=\frac{1}{\beta}\int_0^R dr_1
   \left[
      \ln{r_1}\,e^{-\alpha r_1}
      \left(
         e^{-\beta(R-r_1)}-e^{-\beta(R+r_1)}
      \right)
   \right]
\\[4mm]\displaystyle\hspace{20mm}
=\frac{e^{-\beta R}}{\beta}
\left[
   \int_0^R dr_1\,\ln{r_1}\>e^{-(\alpha-\beta) r_1} -
   \int_0^R dr_1\,\ln{r_1}\>e^{-(\alpha+\beta) r_1}
\right]
\\[4mm]\displaystyle\hspace{20mm}
=\frac{e^{-\beta R}}{\beta}\times
\Biggl[
  \frac{\mbox{E}_1\!\bigl(R(\alpha\!+\!\beta)\bigr)+\ln{(\alpha\!+\!\beta)}
       +\ln{R}\cdot e^{-(\alpha+\beta)R}+\gamma_E}{\alpha+\beta}
\\[4mm]\displaystyle\hspace{38mm}
  -\,\frac{\mbox{E}_1\!\bigl(R(\alpha\!-\!\beta)\bigr)+\ln{(\alpha\!-\!\beta)}
       +\ln{R}\cdot e^{-(\alpha-\beta)R}+\gamma_E}{\alpha-\beta}
\Biggr]
\end{array}
\end{equation}
\end{subequations}

Summing up the two contributions we obtain a final expression
\begin{equation}
\begin{array}{@{}l}
\displaystyle
\Gamma_{\ln{r_1},00}(\alpha,\beta) =
   -\frac{4\pi}{R}\,\frac{e^{-\beta R}\,\gamma_E+e^{-\alpha R}\ln{R}}{\alpha^2-\beta^2}
   +\frac{2\pi}{R\beta}\;
      \frac{e^{\beta R}\,\mbox{E}_1\!\bigl(R(\alpha\!+\!\beta)\bigr)
      +e^{-\beta R}\ln{(\alpha\!+\!\beta)}}{\alpha+\beta}
\\[4mm]\displaystyle\hspace{67.5mm}
   -\frac{2\pi}{R\beta}\>
   \frac{e^{-\beta R}\left[\mbox{E}_1\!\bigl(R(\alpha\!-\!\beta)\bigr)
            +\ln{(\alpha\!-\!\beta)}\right]}{\alpha-\beta}
\end{array}
\end{equation}

To generate other integrals one may use
\begin{equation}
\Gamma_{\ln{r_1},kl}(\alpha,\beta) =
   \left(-\frac{\partial}{\partial\alpha}\right)^k
   \left(-\frac{\partial}{\partial\beta}\right)^l
      \Gamma_{\ln{r_1},00}(\alpha,\beta).
\end{equation}

\clearpage

\begin{center}
{\large\bf
Calculation of the relativistic Bethe logarithm in the two-center problem\\Supplemental Material}\\[5mm]
Vladimir I. Korobov\\
Bogolyubov Laboratory of Theoretical Physics, Joint Institute
for Nuclear Research,\\ Dubna 141980, Russia\\[3mm]
L.~Hilico, and J.-Ph.~Karr\\
Laboratoire Kastler Brossel, UPMC-Paris-6, ENS, CNRS\\
Case 74, 4 place Jussieu, 75005 Paris, France
\end{center}
\vspace{4mm}

In this Supplemental Material, we give the nonrelativistic Bethe logarithm (Table I) and more extensive results for its relativistic corrections (Tables II and III), both for the ground electronic state of $\mbox{H}_2^+$ and $\mbox{He}^+\bar{p}$. These additional data are helpful for computational use to perform more accurately averaging over vibrational wave functions.

\vspace{-2mm}

\section*{Nonrelativistic Bethe logarithm}

\vspace{-2mm}
\begin{longtable}{d@{\hspace{0mm}}d@{\hspace{0mm}}d@{\hspace{3mm}}}
\caption{The nonrelativistic Bethe logarithm, $\beta_{nr}$, for $\mbox{H}_2^+$ and $\mbox{He}^+\bar{p}$.}\\
\hline\hline
 \multicolumn{1}{c}{\vrule width0pt height10pt depth3pt $R$ (in a.u.)}
 & \multicolumn{1}{r}{$Z_1\!=\!1$, $Z_2\!=\!1$}
 & \multicolumn{1}{r}{$Z_1\!=\!2$, $Z_2\!=\!-1$} \\
\hline
\endfirsthead
\caption[]{(continued)}\\
\hline\hline
 \multicolumn{1}{c}{\vrule width0pt height10pt depth3pt $R$ (in a.u.)}
 & \multicolumn{1}{r}{$Z_1\!=\!1$, $Z_2\!=\!1$}
 & \multicolumn{1}{r}{$Z_1\!=\!2$, $Z_2\!=\!-1$} \\
\hline
\endhead
\hline\hline
\endfoot
   0.00  &   4.370422917  &  2.984128556  \\
   0.05  &   3.960959074  &  4.429093491  \\
   0.10  &   3.763208049  &  4.695144972  \\
   0.15  &   3.627181789  &  4.737059502  \\
   0.20  &   3.525244876  &  4.712352798  \\
   0.25  &   3.445386029  &  4.668731134  \\
   0.30  &   3.381060204  &  4.622934909  \\
   0.35  &   3.328241427  &  4.580920363  \\
   0.40  &   3.284256226  &  4.544480350  \\
   0.45  &   3.247231763  &  4.513745134  \\
   0.50  &   3.215803070  &  4.488209081  \\
   0.55  &   3.188944303  &  4.467169500  \\
   0.60  &   3.165865517  &  4.449913692  \\
   0.65  &   3.145946182  &  4.435793560  \\
   0.70  &   3.128690731  &  4.424249332  \\
   0.75  &   3.113697833  &  4.414810709  \\
   0.80  &   3.100638579  &  4.407088769  \\
   0.85  &   3.089240633  &  4.400764849  \\
   0.90  &   3.079276477  &  4.395579241  \\
   0.95  &   3.070554542  &  4.391320901  \\
   1.00  &   3.062912414  &  4.387818567  \\
   1.05  &   3.056211557  &  4.384933303  \\
   1.10  &   3.050333156  &  4.382552371  \\
   1.15  &   3.045174825  &  4.380584212  \\
   1.20  &   3.040647951  &  4.378954395  \\
   1.25  &   3.036675548  &  4.377602338  \\
   1.30  &   3.033190500  &  4.376478663  \\
   1.35  &   3.030134120  &  4.375543070  \\
   1.40  &   3.027454949  &  4.374762617  \\
   1.45  &   3.025107756  &  4.374110337  \\
   1.50  &   3.023052703  &  4.373564122  \\
   1.55  &   3.021254630  &  4.373105821  \\
   1.60  &   3.019682462  &  4.372720508  \\
   1.65  &   3.018308688  &  4.372395891  \\
   1.70  &   3.017108931  &  4.372121836  \\
   1.75  &   3.016061565  &  4.371889968  \\
   1.80  &   3.015147393  &  4.371693363  \\
   1.85  &   3.014349362  &  4.371526283  \\
   1.90  &   3.013652319  &  4.371383970  \\
   1.95  &   3.013042797  &  4.371262469  \\
   2.00  &   3.012508830  &  4.371158489  \\
   2.05  &   3.012039785  &  4.371069288  \\
   2.10  &   3.011626220  &  4.370992576  \\
   2.15  &   3.011259757  &  4.370926439  \\
   2.20  &   3.010932968  &  4.370869276  \\
   2.25  &   3.010639275  &  4.370819742  \\
   2.30  &   3.010372862  &  4.370776708  \\
   2.35  &   3.010128599  &  4.370739223  \\
   2.40  &   3.009901965  &  4.370706488  \\
   2.45  &   3.009688991  &  4.370677824  \\
   2.50  &   3.009486203  &  4.370652660  \\
   2.55  &   3.009290569  &  4.370630511  \\
   2.60  &   3.009099455  &  4.370610966  \\
   2.65  &   3.008910587  &  4.370593673  \\
   2.70  &   3.008722010  &  4.370578334  \\
   2.75  &   3.008532059  &  4.370564694  \\
   2.80  &   3.008339327  &  4.370552535  \\
   2.85  &   3.008142638  &  4.370541670  \\
   2.90  &   3.007941025  &  4.370531937  \\
   2.95  &   3.007733705  &  4.370523199  \\
   3.00  &   3.007520064  &  4.370515335  \\
   3.05  &   3.007299631  &  4.370508243  \\
   3.10  &   3.007072071  &  4.370501832  \\
   3.15  &   3.006837162  &  4.370496026  \\
   3.20  &   3.006594789  &  4.370490756  \\
   3.25  &   3.006344924  &  4.370485963  \\
   3.30  &   3.006087623  &  4.370481595  \\
   3.35  &   3.005823009  &  4.370477609  \\
   3.40  &   3.005551267  &  4.370473962  \\
   3.45  &   3.005272634  &  4.370470622  \\
   3.50  &   3.004987392  &  4.370467557  \\
   3.55  &   3.004695862  &  4.370464739  \\
   3.60  &   3.004398396  &  4.370462145  \\
   3.65  &   3.004095374  &  4.370459753  \\
   3.70  &   3.003787194  &  4.370457545  \\
   3.75  &   3.003474273  &  4.370455503  \\
   3.80  &   3.003157041  &  4.370453612  \\
   3.85  &   3.002835935  &  4.370451859  \\
   3.90  &   3.002511398  &  4.370450231  \\
   3.95  &   3.002183876  &  4.370448719  \\
   4.00  &   3.001853814  &  4.370447311  \\
   4.05  &   3.001521655  &  4.370445999  \\
   4.10  &   3.001187836  &  4.370444776  \\
   4.15  &   3.000852787  &  4.370443634  \\
   4.20  &   3.000516930  &  4.370442567  \\
   4.25  &   3.000180678  &  4.370441568  \\
   4.30  &   2.999844431  &  4.370440633  \\
   4.35  &   2.999508578  &  4.370439757  \\
   4.40  &   2.999173494  &  4.370438934  \\
   4.45  &   2.998839540  &  4.370438162  \\
   4.50  &   2.998507062  &  4.370437437  \\
   4.55  &   2.998176392  &  4.370436754  \\
   4.60  &   2.997847844  &  4.370436112  \\
   4.65  &   2.997521719  &  4.370435506  \\
   4.70  &   2.997198299  &  4.370434936  \\
   4.75  &   2.996877852  &  4.370434397  \\
   4.80  &   2.996560629  &  4.370433889  \\
   4.85  &   2.996246864  &  4.370433409  \\
   4.90  &   2.995936776  &  4.370432955  \\
   4.95  &   2.995630567  &  4.370432525  \\
   5.00  &   2.995328425  &  4.370432118  \\
   5.05  &   2.995030520  &  4.370431733  \\
   5.10  &   2.994737010  &  4.370431368  \\
   5.15  &   2.994448035  &  4.370431021  \\
   5.20  &   2.994163724  &  4.370430693  \\
   5.25  &   2.993884189  &  4.370430380  \\
   5.30  &   2.993609531  &  4.370430084  \\
   5.35  &   2.993339837  &  4.370429802  \\
   5.40  &   2.993075180  &  4.370429533  \\
   5.45  &   2.992815623  &  4.370429278  \\
   5.50  &   2.992561218  &  4.370429035  \\
   5.55  &   2.992312003  &  4.370428803  \\
   5.60  &   2.992068009  &  4.370428583  \\
   5.65  &   2.991829253  &  4.370428372  \\
   5.70  &   2.991595748  &  4.370428171  \\
   5.75  &   2.991367492  &  4.370427980  \\
   5.80  &   2.991144480  &  4.370427797  \\
   5.85  &   2.990926695  &  4.370427622  \\
   5.90  &   2.990714115  &  4.370427455  \\
   5.95  &   2.990506709  &  4.370427295  \\
   6.00  &   2.990304442  &  4.370427142  \\
   6.05  &   2.990107272  &  4.370426996  \\
   6.10  &   2.989915150  &  4.370426856  \\
   6.15  &   2.989728023  &  4.370426722  \\
   6.20  &   2.989545835  &  4.370426593  \\
   6.25  &   2.989368523  &  4.370426470  \\
   6.30  &   2.989196022  &  4.370426352  \\
   6.35  &   2.989028262  &  4.370426239  \\
   6.40  &   2.988865171  &  4.370426130  \\
   6.45  &   2.988706674  &  4.370426025  \\
   6.50  &   2.988552692  &  4.370425925  \\
   6.55  &   2.988403148  &  4.370425829  \\
   6.60  &   2.988257957  &  4.370425736  \\
   6.65  &   2.988117039  &  4.370425647  \\
   6.70  &   2.987980307  &  4.370425562  \\
   6.75  &   2.987847676  &  4.370425479  \\
   6.80  &   2.987719061  &  4.370425400  \\
   6.85  &   2.987594373  &  4.370425324  \\
   6.90  &   2.987473527  &  4.370425251  \\
   6.95  &   2.987356433  &  4.370425180  \\
   7.00  &   2.987243004  &  4.370425112  \\
\end{longtable}

\section*{Relativistic correction to the Bethe logarithm}

\vspace{-2mm}
\begin{longtable}{@{\hspace{5mm}}c@{\hspace{1mm}}d@{\hspace{1mm}}d@{\hspace{4mm}}
               d@{\hspace{1mm}}d@{\hspace{5mm}}d@{\hspace{2mm}}d}
\caption{Relativistic Bethe logarithm for the ground ($1s\sigma_g$) electronic state for $Z_1 = Z_2 = 1$.}\\
\hline\hline
\vrule width0pt height10pt depth4pt
$R$ & \multicolumn{1}{c}{$\beta_1^{(a)}$} & \multicolumn{1}{c}{$\beta_1^{(b)}$} & \multicolumn{1}{c}{$\beta_2$} &
      \multicolumn{1}{c}{$\beta_3$} & \mathcal{L}  & \multicolumn{1}{c}{$\mathcal{L}/N(R)$} \\
\hline
\endfirsthead
\caption[]{(continued)}\\
\hline\hline
\vrule width0pt height10pt depth4pt
$R$ & \multicolumn{1}{c}{$\beta_1^{(a)}$} & \multicolumn{1}{c}{$\beta_1^{(b)}$} & \multicolumn{1}{c}{$\beta_2$} &
      \multicolumn{1}{c}{$\beta_3$} & \mathcal{L}  & \multicolumn{1}{c}{$\mathcal{L}/N(R)$} \\
\hline
\endhead
\hline\hline
\endfoot
\vrule width0pt height10pt depth4pt
0.2 &  199.124 & -362.792 & -718.74  & 500.9   & -381.5   & -9.4774  \\
0.3 &  71.425  & -155.864 & -408.41  & 244.09  & -248.76  & -7.7149  \\
0.4 &  26.866  & -73.891  & -293.55  & 153.814 & -186.76  & -7.0981  \\
0.5 &  8.5247  & -36.4368 & -232.469 & 111.218 & -149.163 & -6.8191  \\
0.6 &  0.3419  & -17.8754 & -192.985 & 86.9905 & -123.529 & -6.6763  \\
0.7 & -3.4173  & -8.19603 & -164.670 & 71.3995 & -104.884 & -6.5980  \\
0.8 & -5.1044  & -2.99381 & -143.164 & 60.5041 & -90.7581 & -6.5534  \\
0.9 & -5.7669  & -0.16341 & -126.255 & 52.4484 & -79.7371 & -6.5273  \\
1.0 & -5.94144 &  1.36042 & -112.653 & 46.2540 & -70.9802 & -6.5145  \\
1.1 & -5.85242 &  2.14699 & -101.528 & 41.3527 & -63.8807 & -6.5084  \\
1.2 & -5.64166 &  2.50981 & -92.3095 & 37.3908 & -58.0506 & -6.5071  \\
1.3 & -5.37875 &  2.62712 & -84.5938 & 34.1346 & -53.2109 & -6.5097  \\
1.4 & -5.10067 &  2.60269 & -78.0750 & 31.4227 & -49.1503 & -6.5145  \\
1.5 & -4.82677 &  2.49792 & -72.5270 & 29.1400 & -45.7159 & -6.5212  \\
1.6 & -4.56684 &  2.34961 & -67.7744 & 27.2013 & -42.7903 & -6.5294  \\
1.7 & -4.32531 &  2.17994 & -63.6799 & 25.5425 & -40.2828 & -6.5387  \\
1.8 & -4.10371 &  2.00222 & -60.1351 & 24.1145 & -38.1221 & -6.5489  \\
1.9 & -3.90197 &  1.82441 & -57.0531 & 22.8786 & -36.2520 & -6.5598  \\
2.0 & -3.71921 &  1.65108 & -54.3637 & 21.8045 & -34.6274 & -6.5712  \\
2.1 & -3.55413 &  1.48479 & -52.0099 & 20.8676 & -33.2117 & -6.5831  \\
2.2 & -3.40532 &  1.32679 & -49.9446 & 20.0480 & -31.9752 & -6.5953  \\
2.3 & -3.27133 &  1.17755 & -48.1290 & 19.3294 & -30.8933 & -6.6077  \\
2.4 & -3.15078 &  1.03708 & -46.5306 & 18.6985 & -29.9458 & -6.6203  \\
2.5 & -3.04237 &  0.90509 & -45.1223 & 18.1441 & -29.1155 & -6.6329  \\
2.6 & -2.94493 &  0.78117 & -43.8811 & 17.6566 & -28.3883 & -6.6456  \\
2.7 & -2.85740 &  0.66481 & -42.7875 & 17.2283 & -27.7518 & -6.6583  \\
2.8 & -2.77881 &  0.55550 & -41.8248 & 16.8522 & -27.1958 & -6.6709  \\
2.9 & -2.70829 &  0.45272 & -40.9786 & 16.5527 & -26.6814 & -6.6758  \\
3.0 & -2.64509 &  0.35594 & -40.2365 & 16.2347 & -26.2909 & -6.6955  \\
3.1 & -2.58852 &  0.26478 & -39.5878 & 15.9839 & -25.9276 & -6.7075  \\
3.2 & -2.53795 &  0.17881 & -39.0229 & 15.7664 & -25.6156 & -6.7192  \\
3.3 & -2.49285 &  0.09764 & -38.5337 & 15.5791 & -25.3498 & -6.7306  \\
3.4 & -2.45270 &  0.02094 & -38.1126 & 15.4188 & -25.1256 & -6.7416  \\
3.5 & -2.41708 & -0.05155 & -37.7533 & 15.2831 & -24.9389 & -6.7522  \\
3.6 & -2.38556 & -0.12013 & -37.4498 & 15.1695 & -24.7860 & -6.7624  \\
3.7 & -2.35780 & -0.18499 & -37.1970 & 15.0761 & -24.6637 & -6.7722  \\
3.8 & -2.33346 & -0.24634 & -36.9899 & 15.0008 & -24.5689 & -6.7815  \\
3.9 & -2.31223 & -0.30433 & -36.8243 & 14.9419 & -24.4990 & -6.7903  \\
4.0 & -2.29385 & -0.35913 & -36.6963 & 14.8980 & -24.4513 & -6.7985  \\
4.1 & -2.27805 & -0.41085 & -36.6028 & 14.8674 & -24.4243 & -6.8065  \\
4.2 & -2.26461 & -0.45960 & -36.5387 & 14.8490 & -24.4139 & -6.8135  \\
4.3 & -2.25331 & -0.50549 & -36.5027 & 14.8414 & -24.4201 & -6.8202  \\
4.4 & -2.24395 & -0.54860 & -36.4913 & 14.8435 & -24.4404 & -6.8264  \\
4.5 & -2.23635 & -0.58902 & -36.5019 & 14.8542 & -24.4730 & -6.8320  \\
4.6 & -2.23034 & -0.62682 & -36.5320 & 14.8726 & -24.5165 & -6.8370  \\
4.7 & -2.22576 & -0.66208 & -36.5793 & 14.8978 & -24.5694 & -6.8416  \\
4.8 & -2.22246 & -0.69487 & -36.6417 & 14.9288 & -24.6303 & -6.8456  \\
4.9 & -2.22031 & -0.72527 & -36.7172 & 14.9649 & -24.6979 & -6.8492  \\
5.0 & -2.21917 & -0.75335 & -36.8039 & 15.0053 & -24.7712 & -6.8523  \\
5.1 & -2.21893 & -0.77921 & -36.9002 & 15.0493 & -24.8490 & -6.8549  \\
5.2 & -2.21949 & -0.80291 & -37.0044 & 15.0964 & -24.9304 & -6.8571  \\
5.3 & -2.22073 & -0.82454 & -37.1151 & 15.1459 & -25.0145 & -6.8588  \\
5.4 & -2.22257 & -0.84421 & -37.2308 & 15.1972 & -25.1004 & -6.8602  \\
5.5 & -2.22492 & -0.86200 & -37.3505 & 15.2499 & -25.1875 & -6.8612  \\
5.6 & -2.22770 & -0.87800 & -37.4729 & 15.3036 & -25.2750 & -6.8619  \\
5.7 & -2.23084 & -0.89231 & -37.5970 & 15.3578 & -25.3624 & -6.8623  \\
5.8 & -2.23427 & -0.90504 & -37.7220 & 15.4121 & -25.4493 & -6.8624  \\
5.9 & -2.23794 & -0.91629 & -37.8470 & 15.4662 & -25.5350 & -6.8623  \\
6.0 & -2.24178 & -0.92615 & -37.9714 & 15.5199 & -25.6194 & -6.8619  \\
6.1 & -2.24576 & -0.93472 & -38.0943 & 15.5729 & -25.7019 & -6.8613  \\
6.2 & -2.24982 & -0.94211 & -38.2154 & 15.6250 & -25.7824 & -6.8606  \\
6.3 & -2.25394 & -0.94839 & -38.3342 & 15.6760 & -25.8606 & -6.8597  \\
6.4 & -2.25808 & -0.95368 & -38.4501 & 15.7256 & -25.9363 & -6.8587  \\
6.5 & -2.26220 & -0.95805 & -38.5630 & 15.7739 & -26.0093 & -6.8576  \\
6.6 & -2.26629 & -0.96158 & -38.6725 & 15.8207 & -26.0797 & -6.8563  \\
6.7 & -2.27032 & -0.96439 & -38.7784 & 15.8658 & -26.1473 & -6.8551  \\
6.8 & -2.27427 & -0.96652 & -38.8806 & 15.9094 & -26.2120 & -6.8537  \\
6.9 & -2.27814 & -0.96805 & -38.9789 & 15.9512 & -26.2739 & -6.8523  \\
7.0 & -2.28190 & -0.96905 & -39.0732 & 15.9913 & -26.3328 & -6.8509  \\
\end{longtable}

\clearpage

\begin{longtable}{@{\hspace{2mm}}c@{\hspace{5mm}}d@{\hspace{6mm}}d@{\hspace{1mm}}d@{\hspace{5mm}}d@{\hspace{1mm}}d}
\caption{Relativistic Bethe logarithm for the ground ($1s\sigma$) electronic state, for $Z_1=2$ and $Z_2=-1$.}\\
\hline\hline
\vrule width0pt height10pt depth4pt
$R$ & \multicolumn{1}{c}{$\beta_1$} & \multicolumn{1}{c}{$\beta_2$} & \multicolumn{1}{c}{$\beta_3$} &
      \mathcal{L}  & \multicolumn{1}{c}{$\mathcal{L}/N(R)$} \\
\hline
\endfirsthead
\caption[]{(continued)}\\
\hline\hline
\vrule width0pt height10pt depth4pt
$R$ & \multicolumn{1}{c}{$\beta_1$} & \multicolumn{1}{c}{$\beta_2$} & \multicolumn{1}{c}{$\beta_3$} &
      \mathcal{L}  & \multicolumn{1}{c}{$\mathcal{L}/N(R)$} \\
\hline
\endhead
\hline\hline
\endfoot
\vrule width0pt height10pt depth4pt
0.10  &   192.1   &   -102.   & -381.08  &  -290.9 &  -7.3871 \\
0.15  &    38.01  &   -288.   &  -81.305 &  -331.2 &  -6.9913 \\
0.20  &    29.65  &   -584.1  &   62.514 &  -491.9 &  -8.6471 \\
0.25  &    17.558 &   -799.9  &  159.373 &  -622.9 &  -9.1887 \\
0.30  &     1.741 &   -998.9  &  236.611 &  -760.5 &  -9.5198 \\
0.35  &   -15.627 &  -1191.9  &  305.525 &  -902.0 &  -9.7143 \\
0.40  &   -33.158 &  -1382.7  &  369.822 & -1046.0 &  -9.8402 \\
0.45  &   -49.903 &  -1566.8  &  430.590 & -1186.1 &  -9.8973 \\
0.50  &   -65.195 &  -1743.3  &  487.854 & -1320.6 &  -9.9213 \\
0.55  &   -78.929 &  -1909.9  &  541.306 & -1447.5 &  -9.9272 \\
0.60  &   -91.000 &  -2064.5  &  590.636 & -1564.8 &  -9.9209 \\
0.65  &  -101.172 &  -2207.3  &  635.664 & -1672.8 &  -9.9137 \\
0.70  &  -109.370 &  -2335.7  &  676.345 & -1768.7 &  -9.8946 \\
0.75  &  -116.075 &  -2451.0  &  712.787 & -1854.2 &  -9.8745 \\
0.80  &  -121.547 &  -2553.7  &  745.209 & -1930.0 &  -9.8546 \\
0.85  &  -125.908 &  -2645.8  &  773.901 & -1997.8 &  -9.8408 \\
0.90  &  -127.846 &  -2727.0  &  799.189 & -2055.6 &  -9.8199 \\
0.95  &  -130.235 &  -2797.5  &  821.408 & -2106.3 &  -9.8013 \\
1.00  &  -131.770 &  -2860.8  &  840.902 & -2151.6 &  -9.7898 \\
1.05  &  -132.958 &  -2916.3  &  857.985 & -2191.2 &  -9.7795 \\
1.10  &  -133.766 &  -2963.8  &  872.905 & -2224.6 &  -9.7653 \\
1.15  &  -134.277 &  -3005.3  &  885.953 & -2253.6 &  -9.7521 \\
1.20  &  -134.563 &  -3041.6  &  897.365 & -2278.7 &  -9.7402 \\
1.25  &  -134.682 &  -3073.4  &  907.349 & -2300.7 &  -9.7297 \\
1.30  &  -134.679 &  -3101.3  &  916.090 & -2319.8 &  -9.7204 \\
1.35  &  -134.586 &  -3125.7  &  923.750 & -2336.5 &  -9.7119 \\
1.40  &  -134.430 &  -3147.1  &  930.468 & -2351.0 &  -9.7042 \\
1.45  &  -134.232 &  -3165.9  &  936.368 & -2363.7 &  -9.6974 \\
1.50  &  -134.006 &  -3182.5  &  941.556 & -2374.9 &  -9.6915 \\
1.55  &  -133.765 &  -3197.1  &  946.122 & -2384.7 &  -9.6861 \\
1.60  &  -133.517 &  -3209.9  &  950.148 & -2393.2 &  -9.6810 \\
1.65  &  -133.268 &  -3221.3  &  953.703 & -2400.8 &  -9.6768 \\
1.70  &  -133.025 &  -3231.3  &  956.847 & -2407.4 &  -9.6727 \\
1.75  &  -132.788 &  -3240.3  &  959.631 & -2413.4 &  -9.6696 \\
1.80  &  -132.561 &  -3248.2  &  962.102 & -2418.6 &  -9.6664 \\
1.85  &  -132.344 &  -3255.2  &  964.299 & -2423.2 &  -9.6635 \\
1.90  &  -132.140 &  -3261.4  &  966.255 & -2427.2 &  -9.6607 \\
1.95  &  -131.947 &  -3267.0  &  968.001 & -2430.9 &  -9.6585 \\
2.00  &  -131.766 &  -3272.0  &  969.562 & -2434.2 &  -9.6565 \\
\end{longtable}


\begin{thebibliography}{99}
\bibitem{Bre12} J.C.J.~Koelemeij, B.~Roth, A.~Wicht, I.~Ernsting, and S.~Schiller, Phys. Rev. Lett. {\bf98}, 173002 (2007); U. Bressel, A. Borodin, J. Shen, M. Hansen, I. Ernsting, and S. Schiller, Phys.\ Rev.\ Lett.\ \textbf{108}, 183003 (2012); J.~Shen, A.~Borodin, M.~Hansen, and S.~Schiller, Phys.\ Rev.~A \textbf{85}, 032519 (2012).

\bibitem{Koel12} J.C.J.~Koelemeij, D.W.E.~Noom , D.~de Jong, M.A.~Haddad, and W.~Ubachs, Appl.\ Phys.\ B\ \textbf{107}, 1075 (2012).

\bibitem{Karr12} J.-Ph.~Karr, A.~Douillet, and L.~Hilico, Appl.\ Phys.\ B\ \textbf{107}, 1043 (2012).

\bibitem{Hori11} M.~Hori, A.~S\'ot\'er, D.~Barna, A.~Dax, R.~Hayano, S.~Friedreich, B.~Juh\'asz, Th.~Pask, E.~Widmann, D.~Horv\'ath, L.~Venturelli, and N.~Zurlo, Nature \ \textbf{475}, 484 (2011).

\bibitem{Kor08} V.I.~Korobov, Phys.\ Rev.~A \textbf{77}, 022509 (2008).

\bibitem{Kor08b} V.I.~Korobov, Phys.\ Rev.~A \textbf{77}, 042506 (2008).

\bibitem{Eides} M.I.~Eides, H.~Grotch, and V.A.~Shelyuto, \emph{Theory of Light Hydrogenic Bound States}, Springer Tracks in Modern Physics \textbf{222}, Springer-Verlag, Berlin, Heidelberg (2007).

\bibitem{SapYen} J.R.~Sapirstein, D.R.~Yennie, in: T.~Kinoshita (Ed.),  {\em Quantum Electrodynamics}, World Scientific, Singapore, 1990.

\bibitem{Pac92} K.~Pachucki, Phys.\ Rev.~A \textbf{46}, 648 (1992); K~Pachucki, Ann.\ Phys.\ (N.Y.) \textbf{226}, 1 (1993).

\bibitem{Jen03} U.D.~Jentschura, E.-O.~Le Bigot, P.J.~Mohr, P.~Indelicato, and G.~Soff, Phys.\ Rev.\ Lett.\ \textbf{90}, 163001 (2003).

\bibitem{PRL05} A.~Czarnecki, U.D.~Jentschura, and K.~Pachucki, Phys.\ Rev.\ Lett.\ \textbf{95}, 180404 (2005).

\bibitem{JCP05} U.D.~Jentschura, A.~Czarnecki, and K.~Pachucki, Phys.\ Rev.~A \textbf{72}, 062102 (2005).

\bibitem{CasLep} W.E.~Caswell and J.P.~Lepage, Phys.\ Lett.~B \textbf{167}, 437 (1986).

\bibitem{Kin96} T.~Kinoshita and M.~Nio, Phys.\ Rev.~D \textbf{53}, 4909 (1996).

\bibitem{Pac98} K.~Pachucki, J.~Phys.~B \textbf{31}, 3547 (1998).

\bibitem{Bet47} H.A.~Bethe, Phys.\ Rev.\ \textbf{72}, 339 (1947).

\bibitem{BS} H.A.~Bethe and E.E.~Salpeter, {\em Quantum mechanics of one-- and two--electron atoms}, Plenum Publishing Co., New York, 1977.

\bibitem{HFS09} V.I.~Korobov, L.~Hilico, and J.-Ph.~Karr, Phys.\ Rev.~A \textbf{79}, 012501 (2009).

\bibitem{Schwartz} C.~Schwartz, Phys.\ Rev.\ \textbf{123}, 1700 (1961).

\bibitem{Kor12} V.I.~Korobov, Phys.\ Rev.~A \textbf{85}, 042514 (2012).

\bibitem{Gav70} M.~Gavrila and A.~Costescu, Phys.\ Rev.~A \textbf{2}, 1752 (1970).

\bibitem{Tso06} Ts.~Tsogbayar and V.I.~Korobov, J.~Chem.\ Phys.\ \textbf{125}, 024308 (2006).

\bibitem{Kor00} V.I.~Korobov, Phys.\ Rev.~A {\bf 61}, 064503 (2000).

\bibitem{Kor07} V.I.~Korobov and Ts.~Tsogbayar, J.~Phys.~B \textbf{80}, 2661 (2007).

\bibitem{Kolos} R.~Bukowski, B.~Jeziorski, R.~Moszy\'nski, and W.~Ko\l os, Int.\ J.~Quantum Chem.\ \textbf{42}, 287 (1992).

\bibitem{Supp} See Supplemental Material below for complete Tables of the numerical data.

\end{thebibliography}
\end{document}